\documentclass[showpacs,]{revtex4}
\usepackage{bm}
\usepackage{amsmath}
\usepackage{epsfig}
\usepackage{graphicx}
\usepackage{color}
\newcommand{\bfvec}[1]{\hbox{\boldmath$#1$\unboldmath}}

\newcommand{\fslash}[1]{{#1\hspace*{-0.5em}\slash}}
\begin{document}

\title{Relativistic Chiral Hartree-Fock description of nuclear matter 
with constraints from nucleon structure and confinement }
\author{E. Massot, G. Chanfray} 
\affiliation{IPN Lyon, Universit\'e de Lyon, Univ.  Lyon 1, 
 CNRS/IN2P3, UMR5822, F-69622 Villeurbanne Cedex}
\begin{abstract}
We present a relativistic chiral effective theory for symmetric and asymmetric nuclear matter taken in  the Hartree-Fock scheme. 
The nuclear binding is insured by a background chiral invariant scalar field associated with the radial fluctuations 
of the chiral quark condensate. Nuclear matter saturation is obtained once the scalar response of the nucleon generating three-body repulsive forces 
is incorporated. For these  parameters related to the scalar sector and quark confinement mechanism inside the nucleon we make use 
of  an analysis of lattice results  on the 
nucleon mass evolution with the quark mass. The other parameters are constrained as most as possible by standard hadron and nuclear phenomenology. Special attention is paid to the treatment of the propagation of the scalar fluctuations. The rearrangement terms associated with in-medium modified mass and coupling constants are explicitly included to satisfy the Hugenholtz -Van Hove theorem.
We point out the important role of the tensor piece of the rho exchange Fock term to reproduce the asymmetry energy of nuclear matter. We also discuss the isospin dependence of the Landau nucleon  effective mass.
\end{abstract}

\maketitle
%%%%%%%%%%%%%%%%%%%%%%%%%%%%%%%%%%%%%%%%%%%%%%%%%%%%%%%%%%%%%%%%%%%%%%%%%%%%%ש
\section{Introduction}\label{Intro}
%%%%%%%%%%%%%%%%%%%%%%%%%%%%%%%%%%%%%%%%%%%%%%%%%%%%%%%%%%%%%%%%%%%%%%%%%%%%%%%%%
A fundamental question of present day nuclear physics is to relate low energy non perturbative QCD and in first rank chiral symmetry and confinement to the very rich structure  of the nuclear many-body problem. However  it is presently hopeless to  derive the observed nuclear properties from the underlying QCD and a more modest ambition is to put some constraints on the modelling of nuclear matter properties not only from hadronic phenomenology but also from lattice QCD data. This may also constitute a starting point to elucidate an old and central  question of strong interaction physics, namely   the interrelation between the many-body effects governing the equation of state of nuclear matter and the  nucleon substructure response  to the nuclear  environmment.\\

A first attempt to go beyond the standard non relativistic treatment of nuclear matter is the relativistic mean field approach initiated by Walecka and collaborators \cite{SW86}. In this framework the nucleons move in an attractive background scalar field and in a repulsive vector background field. This provides a very economical saturation mechanism and a spectacular well known success is the correct magnitude of the spin-orbit potential since the large vector and scalar fields contribute to it in an  additive way. Another successful modern attempt  is based on in-medium chiral perturbation theory where a pion loop expansion is performed on top of scalar and vector background scalar fields using a density functional formulation \cite{FKVW06}. Now the question of the very nature of these background fields has to be elucidated or said differently it is highly desirable to clarify their  relationship with the QCD condensates and in particular the chiral quark condensate.  

To address this question we take the point of view that the effective theory has to be  formulated, as a starting point,  in term of the fields associated with the fluctuations of the chiral quark condensate parametrized in a matrix form $W=\sigma + i\vec\tau\cdot\vec\pi$. 
The sigma and the pion, associated with the amplitude and phase fluctuations of this condensate  are promoted to the rank of effective degrees of freedom. Their dynamics are governed by an  effective potential, $V(\sigma, \vec{\pi})$, having a  typical mexican hat shape associated with a broken (chiral) symmetry of the QCD vacuum. Explicit construction of such an effective theory for the description of nuclear matter can be performed for instance within the NJL model \cite{BT01}.

As proposed in a previous paper \cite{CEG02} an alternative and very convenient formulation of the resulting sigma model is obtained by going from cartesian to polar coordinates  {\it i.e.}, going from a linear to a non linear representation,  according to~:
$W=\sigma\, + \,i\vec\tau\cdot\vec\pi=S\,U=(f_\pi\,+\,s)\,exp\left({i\vec\tau\cdot\vec\varphi_\pi/ f_\pi}\right)$.
The new pion field $\vec\varphi_\pi$ corresponds to an orthoradial soft mode which is automatically massless (in the absence of explicit chiral symmetry breaking) since it is associated with rotations on the chiral circle without cost of energy. 
The new sigma meson field, $S$, which is a chiral invariant,  describes a radial mode associated with the fluctuations of the ``chiral radius'' around its vacuum expectation value, $f_\pi$. It can be associated with the ordinary sigma meson 
which gets a very large width from its strong decay into two pions. Since it has derivative couplings to the pion field, it decouples 
from low energy pions whose dynamics is described by chiral perturbation theory. The evolution of the expectation value of $S$ is related to the non pionic contribution to the in-medium chiral condensate \cite{CEG02,EC07}. This expectation value plays the role of a chiral order parameter around the minimum of the effective potential and  the medium can be seen as a shifted vacuum. With increasing density, its fluctuations $s=S-f_\pi$ are associated with the progressive shrinking of the chiral circle and it governs the evolution of the nucleon mass. Here comes our main physical assumption proposed  in ref. \cite{CEG02}. We identify this chiral invariant $s$ field with the sigma meson of nuclear physics and relativistic theories of the Walecka type, or, said differently, with the background attractive scalar field at the origin of the nuclear binding. This also gives a plausible answer to the long-standing problem of the chiral status of Walecka theories.  

One motivation of the present work is to study in some details whether this hypothesis yields a viable  description of nuclear matter. It is nevertheless well known that in such chiral theories, independently of the details of the modelling, tadpole diagrams associated with the mexican hat potential automatically generate attractive three-body forces destroying saturation \cite{KM74,BT01}. 
The origin of this failure can be attributed to the neglect of the effect of nucleon substructure linked to the confinement mechanism as already pointed out in some of our previous works \cite{CE05,CE07,EC07}. 

Our article is organized as follows. The second section, which is in some sense  a brief summary of our previous works, is devoted to the constraints brought by lattice data for the description of the nucleon and nuclear matter. In section \ref{CHI} we present the chiral lagrangian and section \ref{HAM} is devoted to the construction of the hamiltonian in the static approximation; we also give a detailed description of the treatment of the propagation of the in-medium modified scalar field. The Hartree-Fock approach including rearrangement terms is presented in section \ref{HAR} and is applied to the case of infinite matter in section \ref{INF}. Finally in section \ref{RES} numerical results are given and the results discussed.
%%%%%%%%%%%%%%%%%%%%%%%%%%%%%%%%%%%%%%%%%%%%%%%%%%%%%%%%%%%%%%%%%%%%%%%%%%%%%
\section{Constraints on the chiral effective theory from QCD susceptibilities}\label{CON}
%%%%%%%%%%%%%%%%%%%%%%%%%%%%%%%%%%%%%%%%%%%%%%%%%%%%%%%%%%%%%%%%%%%%%%%%%%%%%%%%%%%% 
\subsection{Tests of the effective theory with a chiral invariant scalar field}\label{TES} 
Once the appropriate couplings of the chiral fields to the baryons are introduced one can build an effective lagrangian to describe nuclear matter. Vector mesons ($\omega$ and $\rho$) must be also included to get the needed short range repulsion and asymmetry properties (see sections \ref{CHI} and \ref{RES}). At the Hartree level, the pion and the rho do not contribute for symmetric nuclear matter whose energy density written as a function of the order parameter $\bar s=\langle s \rangle$ is~:
\begin{equation}
{E_0\over V}=\varepsilon_0=\int\,{4\,d^3 p\over (2\pi)^3} \,\Theta(p_F - p)\,E^*_p(\bar s)
\,+\,V(\bar s)\,+\,{g^2_\omega\over 2\, m_\omega^2}\,\rho^2.\label{HARTREE}
\end{equation}
 $E^*_p(\bar s)=\sqrt{p^2\,+\,M^{*2}_N(\bar s)}$ is the energy of an 
effective nucleon with the effective Dirac mass $M^*_N(\bar s)=M_N +g_S\,\bar s$. $g_S$ is the scalar coupling constant of the model; in the pure linear sigma model it is $g_S=M_N/f_\pi$. The effective potential $V(\sigma, \vec{\pi})$ when reexpressed in term of the new polar representation has the typical form~:
$$V(s)=\frac{1}{2} m^{2}_{\sigma}\,\left(s^2\,+\,\frac{1}{2}\,\frac{s^3}{f_\pi}\,+...\right).$$
$\bar s$ is obtained by minimization of the energy density and is given at low density by~: $\bar s\approx-(g_S/ m_\sigma^2)\,\rho_S$. Its negative value is at the origin of the binding but the presence of the $s^3$ term (tadpole) has very important consequences as already mentioned in the introduction. This tadpole is at the origin of the chiral dropping \cite{HKS99} of the sigma mass $\Delta m^{*}_{\sigma} \simeq  -(3\,g_S /2  f_{\pi})\rho_S$ (a $\simeq 30\%$ effect at $\rho_0$) and generates an attractive three-body force which makes nuclear matter collapse and destroys the Walecka saturation mechanism. Hence the chiral theory does not pass the nuclear matter stability test. 

\smallskip\noindent
This failure, which is in fact a long-standing problem \cite{KM74,BT01}, is maybe not so surprising since  the theory, as it is, also fails to describe some nucleon structure aspects as discussed below. The nucleon mass, as well as other intrinsic properties of the nucleon (sigma term, chiral susceptibilities),
are QCD quantities which are in principle obtainable from lattice simulations. The problem is that lattice calculations of this kind are not feasible for quark masses smaller than $50$ MeV, or equivalently pion mass smaller than $400$ MeV, using the GOR relation. Hence one needs a technics to extrapolate the lattice data to the physical region. The difficulty of the extrapolation is linked to the non analytical behaviour of the nucleon mass as a function of $m_q$ (or equivalently $m^2_\pi$) which comes from the pion cloud contribution. The idea of Thomas {\it et al} \cite{TGLY04} was to separate the pion cloud self-energy, $\Sigma_{\pi}(m_{\pi}, \Lambda)$, from the rest of the nucleon mass and to calculate it in a chiral model with  one adjustable cutoff parameter $\Lambda$. They expanded the remaining part  in terms of $m^2_{\pi}$  as follows~:
\begin {equation}
M_N(m^{2}_{\pi}) = 
a_{0}\,+\,a_{2}\,m^{2}_{\pi}\, +\,a_{4}\,m^{4}_{\pi}\,+\,\Sigma_{\pi}(m_{\pi}, \Lambda)\,.\label{LATTICE}
\end{equation}
At this point it is important to stress that the above expansion is in reality an expansion in terms of the current quark mass $m_q$ which is the genuine parameter occuring in the lattice calculation. The pion mass appearing in eq. (\ref{LATTICE}) is just the pion mass deduced from the quark mass assuming the GOR relation $m^2_{\pi}=-2\,m_q\,\langle\bar q q\rangle_{vac}/f^2_{\pi}$. This reparametrization has been adopted only for convenience. 
The best fit value of the parameter $a_{4}$   shows little sensitivity to the shape of the form factor, with a value 
$a_4 \simeq- 0.5\, \mathrm{GeV}^{-3} $ while $a_2 \simeq 1.5\,\mathrm{GeV}^{-1}$ (see ref. \cite{TGLY04}). The small value 
of $a_4$ reflects the fact that the non pionic contribution to the nucleon mass is almost linear in $m^2_{\pi}$ ({\it i.e.}, in $m_q$).
Taking successive derivatives of $M_N$ with respect to $m^2_{\pi}$ ({\it i.e.}, to $m_q$), it is possible to obtain some fundamental chiral properties of the nucleon,  namely the pion-nucleon sigma term and the scalar susceptibility of the nucleon. The non pionic pieces of these  quantities are given by~:
\begin{equation}
 \sigma_N^{non-pion} =m_q\, \frac{\partial M_N^{non-pion}}{\partial m_q}\simeq m^2_{\pi} {\partial M\over \partial m^2_{\pi }}
=a_2 \,m^2_{\pi}\, + \,2\,a_4 \, m^4_{\pi}\simeq 29\, \mathrm{MeV}\,.
\end{equation}
\begin{equation}
 \chi_{NS}^{non- pion}=\frac{\partial\left(\sigma_N^{non-pion}/2\,m_q\right)}{\partial m_q}\simeq 2{\langle\bar q q\rangle_{vac}^2 \over f^4_{\pi} }
 {\partial~~\over \partial m^2_{\pi }}\left({\sigma_N^{non-pion} \over m^2_{\pi }}\right)=  
 {\langle\bar q q\rangle_{vac}^2 \over f^4_{\pi} }\,4\,a_{4}\,.
\end{equation}
In the above equations the first equalities correspond to the definitions, the second equalities make use of the GOR relation and the last ones 
come from the lattice QCD analysis. With  typical  cutoff used in this analysis, $\Lambda\simeq 1$ GeV, which  yields $\sigma_N^{(\pi)}\simeq20$ MeV, the total value of the sigma term is $\sigma_N\simeq 50 $ MeV, a quite satisfactory result in view of the most recent analysis.
It is interesting to compare what comes out from the lattice approach with our chiral effective model. At this stage the only non pionic contribution 
to the nucleon mass comes from the scalar field, or more microscopically the nucleon mass entirely comes from the chiral condensate 
since the nucleon is just made of three constituent quarks with mass $M_Q=g\langle S \rangle_{vac}=g f_\pi\simeq 350$ MeV. Hence the results for the non pionic sigma term and scalar susceptibility are identical to those of the linear sigma model~:
\begin{equation}
\sigma_N^{(\sigma)}=f_\pi\,g_S\,\frac{m^2_\pi}{m^2_\sigma},\qquad
\chi^{(\sigma)}_{NS}=-2\,\frac{\left\langle \bar q q\right\rangle^2_{vac}}{f_\pi^3}\,
\frac{3\,g_S}{m^4_\sigma}.	
\end{equation}
The identification of   $\sigma_N^{non-pion}$ with $\sigma_N^{(\sigma)}$ of our model
fixes the sigma mass to a value  $m_\sigma=800$ MeV, close to the one $\simeq750$ MeV that we have used 
in a previous article \cite{CE05}. As it is the ratio $g_S/{m^2_\sigma}$ which is thus determined this 
value of $m_\sigma$ is associated with the coupling constant of the linear sigma model 
$g_S=M_N/f_\pi=10$. Lowering $g_S$ reduces $m_\sigma$. Similarly, the identification  of $\chi^{(\sigma)}_{NS}$
with the lattice expression provides a model value for $a_4$. The numerical result is $a_4^{(\sigma)}= -3.4\,\hbox{GeV}^{-3}$
while the value obtained in the expansion is only   $- 0.5\,\hbox{GeV}^{-3}$. 
  
%%%%%%%%%%%%%%%%%%%%%%%%%%%%%%%%%%%%%%%%%%%%%%%%%%%%%%%%%%%%%%%%%%%
\subsection{Nucleon structure effects and  confinement mechanism} 
%%%%%%%%%%%%%%%%%%%%%%%%%%%%%%%%%%%%%%%%%%%%%%%%%%%%%%%%%%%%%%%%%%%%%
The net conclusion of the above discussion is that the model as such fails to pass the QCD test since the $a_4$ coefficient is much larger in the chiral model than the one extracted from the lattice data analysis. 
In fact this is to be expected and even gratifying because it also fails the nuclear physics test as discussed in subsection \ref{TES}. We will see that these two important failures may have a common origin. Indeed  an important effect is  missing, namely  the  scalar response of the nucleon,  
$ \kappa_{NS}=\partial^2 M_N/\partial s^2$, to the scalar 
nuclear field, which is the basis of the quark-meson coupling model (QMC) introduced in ref. \cite{G88}. The physical reason is very easy to understand: the nucleons are quite large composite systems of quarks and gluons and they should respond to the nuclear environment, {\it i.e.},
to the background nuclear scalar  fields. This response originates from the quark wave function modification  in the nuclear field 
and will obviously depend on the confinement mechanism. This confinement effect is expected to generate a positive scalar response 
$\kappa_{NS}$, {\it i.e.,} it opposes an increase of the scalar field, a feature   confirmed by the lattice analysis (see below).
This polarization of the nucleon is accounted for by the phenomenological introduction
of the scalar nucleon response, $\kappa_{NS}$, in  the nucleon mass evolution as follows~:
\begin{equation}
	M_N(s)=M_N\,+\,g_S\,s\,+\,\frac{1}{2}\,\kappa_{NS}\,s^2\,+\,....\label{NUCLEONMASS}
\end{equation}
This constitutes the only change in the expression of the energy density (eq. \ref{HARTREE}) but this has numerous consequences. The effective scalar coupling constant drops with increasing density but  the sigma mass gets stabilized~:
\begin{equation}
g^*_S(\bar s)={\partial M^*_N\over\partial\bar s}={M_N\over f_{\pi}}\,+\,\kappa_{NS}\, \bar s,\quad
m^{*2}_\sigma =\frac{\partial^2 \varepsilon}{\partial\bar s^2}\simeq
m^{2}_{\sigma}- ( {3\,g_S \over f_{\pi}} -  \, \kappa_{NS})\,\rho_S .
\end{equation}
The non-pionic  contribution to the nucleon susceptibility is modified, as well \cite{CE07}~:
\begin{equation}
 \chi_{NS}^{(\sigma)}= -2{\langle\bar q q\rangle_{vac}^2 \over f^2_{\pi} }\,
 \left({1\over m^{*2}_\sigma}\,-\,{1\over m^{2}_\sigma}\right)\,{1\over\rho}=
-2{\langle\bar q q\rangle_{vac}^2 \over f^2_{\pi} }\,{1\over m^{4}_\sigma}\,
\left({3\,g_S\over f_\pi}\,-\,\kappa_{NS}\right)\,. 
\end{equation}
We see that the effect of confinement ($\kappa_{NS}$) is to compensate  the pure scalar term. Again comparing with the lattice expression one gets 
a model value for the $a_4$ parameter~:
\begin{equation}
a_4= -{ a_2^2 \over 2\,M}( 3\,-\,2\,C).
\end{equation}
where $C$ is the dimensionless parameter $ C= \left({f^2_{\pi}/  2\,M} \right) \kappa_{NS} $. Numerically 
$a_4=- 0.5 \,GeV^{-3}$ gives $ C =+1.25$, implying a large cancellation.  As discussed in ref.  \cite{CE05,CE07,EC07} such a significant scalar response will generate other repulsive forces which restore the saturation mechanism. At this point it is important to come again to the underlying physical picture
implying that the nucleon mass originates both from the coupling to condensate and from confinement. In the original formulation of the quark coupling model, nuclear matter is represented as a collection of (MIT) bags seen as bubbles of perturbative vacuum in which quarks are confined. Thus 
in such a picture the mesons should not appear inside the bag  and should  not couple to quarks  as in the true non perturbative  QCD vacuum. Consequently the bag picture is at best an effective realisation of confinement which must not to be taken too literally. Indeed, QCD lattice simulations strongly suggest that a more realistic picture is closer to a $Y$ shaped color string (confinement aspect) attached to quarks
\cite{B05}. Outside this relatively thin string one has the ordinary non perturbative QCD vacuum possessing a chiral condensate from which the quarks get their constituent mass.
%%%%%%%%%%%%%%%%%%%%%%%%%%%%%%%%%%%%%%%%%%%%%%%%%%%%%%%%%%%%%%%%%%%%%%%%%%%%%ש
\section{The chiral Lagrangian}\label{CHI}
%%%%%%%%%%%%%%%%%%%%%%%%%%%%%%%%%%%%%%%%%%%%%%%%%%%%%%%%%%%%%%%%%%%%%%%%%%%%%%%%%
To get a more complete description of symmetric and asymmetric nuclear matter we complete the  lagrangian used in ref. \cite{CEG02} essentially by the introduction of the  rho meson. We also consider  the possibility of  incorporating the scalar isovector $\delta$ meson. Written with obvious notations, it has the form
\begin{equation}
{\cal L}=\bar\Psi\,i\gamma^\mu\partial_\mu\Psi\,+\,
{\cal L}_s\,+\,{\cal L}_\omega\,+\,{\cal L}_\rho\,+\,{\cal L}_\delta\,+\,{\cal L}_\pi
\end{equation}
with
\begin{eqnarray}
{\cal L}_s &=& -M_N(s)\bar\Psi\Psi\,-\,V(s)\,+\,\frac{1}{2} \partial^\mu s\partial_\mu s\nonumber\\	
{\cal L}_\omega &=& - g_\omega\,\omega_\mu\,\bar\Psi\gamma^\mu\Psi\,+\,\frac{1}{2}\,m^2_\omega\,\omega^\mu\omega_\mu
\,-\,\frac{1}{4} \,F^{\mu\nu}F_{\mu\nu}\nonumber\\	
{\cal L}_\rho &=&- g_\rho\,\rho_{a\mu}\,\bar\Psi\gamma^\mu \tau_a\Psi
\,-\,g_\rho\frac{\kappa_\rho}{2\,M_N}\,\partial_\nu \rho_{a\mu}\,\Psi\bar\sigma^{\mu\nu}_{}\tau_a\Psi
\,+\,\frac{1}{2}\,m^2_\rho\,\rho_{a\mu}\rho^{\mu}_{a}
\,-\,\frac{1}{4} \,G_a^{\mu\nu}G_{a\mu\nu}\nonumber\\	
{\cal L}_\delta &=&- g_\delta\,\delta_a\,\bar\Psi\tau_a\Psi\,-\,\frac{1}{2}\,m^2_\delta\,\delta^2
\,+\,\frac{1}{2} \partial^\mu \delta\partial_\mu \delta\nonumber\\	
{\cal L}_\pi &=& \frac{g_A}{2\,f_\pi}\,\partial_\mu\varphi_{a\pi}\bar\Psi\gamma^\mu\gamma^5\tau_a\Psi
-\,\frac{1}{2}\,m^2_{\pi}\varphi_{a\pi}^2
\,+\,\frac{1}{2}\, \partial^\mu\varphi_{a\pi}\partial_\mu \varphi_{a\pi}.
\end{eqnarray}
As discussed previously the form of $M_N(s)$ (eq. \ref{NUCLEONMASS}) reflects the internal nucleon structure and contains a quadratic term involving the scalar response of the nucleon which is constrained by lattice data. However the nucleon mass may very well have higher order derivatives with respect to the scalar field. In practice, as in our previous works \cite{CE05,CE07}, we introduce a cubic term~: 
\begin{equation}
	M_N(s)=M_N\,+\,g_S\,s\,+\,\frac{1}{2}\,\kappa_{NS}\,\left(s^2\,+\,\frac{s^3}{3\,f_\pi}\right).\label{NUCLEONMASSCORR}
	\end{equation}
Hence the scalar susceptibility becomes density dependent~
\begin{equation}
\tilde\kappa_{NS}(s)={\partial^2M_N\over\partial s^2}=
\kappa_{NS}\left(1\,+\,{s\over  f_\pi}\right)
\end{equation} 
and vanishes at full restoration, $\bar s=-f_\pi$, where $\bar s$ is the expectation value of the $s$ field. 
Hidden in the above Lagrangian is the explicit chiral symmetry breaking piece
\begin{equation}
{\cal L}_{\chi SB}=c\,\sigma=-\frac{c}{2}\,Tr (f_\pi\,+\,s)\,exp\left({i\vec\tau\cdot\vec\varphi_\pi/ f_\pi}\right)
\simeq \,c\,s\,-\,\frac{c}{2\,f_\pi}\,\varphi_\pi^2	
\end{equation}
which generates the pion mass term with the identification $c=f_\pi\,m_\pi^2$. It is thus implicit that neglecting the higher order terms in the exponent,  the self-interactions of the pions are omitted. Notice that the only meson having a self-interacting potential $V(s)$ is the scalar meson $s$. We take it in practice as in the linear sigma model with the inclusion of the explicit chiral symmetry breaking piece~:
\begin{eqnarray}
	V(s)&=& {\lambda\over 4}\,\big((f_\pi\,+\,s)^2\,-\,v^2\big)^2\,-\,f_\pi m_\pi^2\,s\nonumber\\
	&\equiv &\frac{m^{2}_{\sigma }}{2}\,s ^{2}\,+\,
\frac{m^{2}_{\sigma}\,-\,m^{2}_{\pi }}{2\,f_\pi}\,s^3\,+\,
\frac{m^{2}_{\sigma}\,-\,m^{2}_{\pi }}{8\,f_\pi^2}\,s^4.
\end{eqnarray}
The other parameters ($g_\omega, g_\rho, \kappa_{\rho}, g_A$ and the meson masses) will be fixed 
as most as possible by hadron phenomenology. It is in principle also possible to calculate or at least to constrain these parameters in an underlying NJL model. 
One specific comment is in order for the tensor coupling of vector mesons. The pure Vector Dominance  picture (VDM) 
implies the identification of $\kappa_{\rho}$ with the anomalous part of the isovector magnetic moment of the nucleon, {\it i.e.}, $\kappa_{\rho}=3.7$. However pion-nucleon scattering data \cite{HP75} suggest $\kappa_{\rho}=6.6$ (strong rho scenario).
We will come to this point later on in the discussion of the results (section \ref{RES}). The omega meson should also possess a tensor coupling but, according to VDM the 
corresponding anomalous isoscalar magnetic moment is $\kappa_{\omega}=0.13$. Since it is very small we neglect it here. For completeness we also add a delta meson which may generate a splitting between the proton and neutron masses but with the chosen coupling constant, $g_\delta=1$, its influence is in practice negligible.
%%%%%%%%%%%%%%%%%%%%%%%%%%%%%%%%%%%%%%%%%%%%%%%%%%%%%%%%%%%%%%%%%%%%%%%%%
\section{Construction of the hamiltonian}\label{HAM}
%%%%%%%%%%%%%%%%%%%%%%%%%%%%%%%%%%%%%%%%%%%%%%%%%%%%%%%%%%%%%%%%%
The conjuguate momenta of the various mesonic fields are~:
\begin{eqnarray}
& &\Pi_s= \partial_0 s,\qquad \Pi_{j\omega}=F^{0j},\qquad\Pi_{a\delta}= \partial_0 \delta_a\,\nonumber\\
& &\Pi_{aj\rho}=G^{a}_{0j}\,+\,g_\rho\frac{\kappa_\rho}{2\,M_N}\,
\bar\Psi\sigma^{j0}\tau_a\Psi,\qquad \Pi_{aj}=\partial_0\varphi_{a\pi}\,-\,\frac{g_A}{2\,f_\pi}\,\bar\Psi\gamma^5\gamma^0\tau_a\Psi .
\end{eqnarray}
The hamiltonian  is obtained  by the usual generalized Legendre transformation with the result~: 
\begin{equation}
H=\int d{\bf r}\,\bigg[\bar\Psi\left(-i\vec{\gamma}\cdot\vec{\nabla}\right)\Psi\,+\,H_s\,+\,H_\omega\,+\,H_\rho
\,+\,H_\delta\,+\,H_\pi\bigg]	
\end{equation}
with~:
\begin{eqnarray}
H_s &=& \int d{\bf r}\,\bigg[M_N(s)\,\bar\Psi\Psi\,+\,\frac{1}{2}\left(\Pi^{2}_{s}\,+(\vec{\nabla}s)^2\right)\,+\,V(s)\bigg]\nonumber\\
H_\omega &=& \int d{\bf r}\,\bigg[g_\omega\,\omega_\mu\,\bar\Psi\gamma^\mu\Psi\,+\,
\frac{1}{2}\,\bigg((\vec\Pi_\omega)^2\,-\, m^{2}_{\omega}\,\omega^\mu\omega_\mu\,+\,\vec{\nabla}\omega^j\cdot \vec{\nabla}\omega^j
\,-\,(\vec{\nabla}\cdot\vec\omega)^2\bigg)\,-\,\vec{\Pi}_\omega\cdot\vec{\nabla}\omega_0\bigg]\nonumber\\
H_\rho &=& \int d{\bf r}\,\bigg[g_\rho\,\rho_{a\mu}\,\bar\Psi\gamma^\mu\tau_a\Psi\,+\,
\,g_\rho\frac{\kappa_\rho}{2\,M_N}\,\bigg(\partial_j \rho_{ai}\,\bar\Psi\sigma^{ij}\tau_a\Psi
\,-\,\Pi_{ai}\,\bar\Psi\sigma^{i0}\tau_a\Psi\bigg)\nonumber\\
& &\, +\,\frac{1}{2}\,
\left(g_\rho\frac{\kappa_\rho}{2\,M_N}\right)^2\,\left(\bar\Psi\sigma^{i0}\tau_a\Psi\right)^2\nonumber\\
& & \,+\,\frac{1}{2}\,\bigg((\vec\Pi_{a\rho})^2\,-\, m^{2}_{\rho}\,\rho_a^\mu\rho_{a\mu}\,+\,
\vec{\nabla}\rho_a^j\cdot \vec{\nabla}\rho_a^j
\,-\,(\vec{\nabla}\cdot\vec\rho_a)^2\bigg)\,-\,\vec{\Pi}_{a\rho}\cdot\vec{\nabla}\rho_{a0}\bigg]\nonumber\\
H_\delta &=& \int d{\bf r}\,\bigg[g_\delta \,\delta_a\,\bar\Psi\tau_a\Psi\,+\,
\frac{1}{2}\left(\Pi^{2}_{a\delta}\,+(\vec{\nabla}\delta_a)^2\,+\,m^{2}_{\delta}\delta^{2}_{a}\right)\bigg]\nonumber\\
H_\pi &=& 	\int d{\bf r}\,\bigg[\frac{g_A}{2\,f_\pi}\,\bigg(\vec{\nabla}\varphi_{a\pi}\cdot\bar\Psi\gamma^5\vec\gamma\tau_a\Psi\,+\,
\vec\Pi_{a\pi}\,\bar\Psi\gamma^5\gamma^0\tau_a\Psi\bigg)\,+\,
\left(\frac{g_A}{2\,f_\pi}\right)^2\,\left(\bar\Psi\gamma^5\gamma^0\tau_a\Psi\right)^2\nonumber\\
& &\,+\,\frac{1}{2}\left(\Pi^{2}_{a\pi}\,+(\vec{\nabla}\varphi_{a\pi})^2\,+\,m^{2}_{\pi}\varphi^{2}_{a\pi}\right)\bigg].
\end{eqnarray}

%%%%%%%%%%%%%%%%%%%%%%%%%%%%%%%%%%%%%%%%%%%%%%%%%%%%%%%%%%%%%%%%%%%%%%%%%%%
\subsection{Static approximation}
%%%%%%%%%%%%%%%%%%%%%%%%%%%%%%%%%%%%%%%%%%%%%%%%%%%%%%%%%%%%%%%%%%%%%%%%%%%%%%%
We will first formulate the Hartree-Fock approach in the static approximation, {\it i.e.}, neglecting the retardation effects. In
such a case the conjuguate momenta are~: 
$$\Pi_s= 0, \vec\Pi_\omega=\vec\nabla\omega^0, \Pi_{a\delta}=0,
\Pi_{aj\rho}=\nabla_j\rho_a^0\,+\,g_\rho\frac{\kappa_\rho}{2\,M_N}\,
\bar\Psi\sigma^{j0}\tau_a\Psi,\, \Pi_{aj}=-\frac{g_A}{2\,f_\pi}\,\bar\Psi\gamma^5\gamma^0\tau_a\Psi.$$

The various pieces of the hamiltonian  simplify according to~:
 \begin{eqnarray}
H_s^{static} &=& \int d{\bf r}\,\bigg[M_N(s)\,\bar\Psi\Psi\,+\,\frac{1}{2}(\vec{\nabla}s)^2\,+\,V(s)\bigg]\nonumber\\
H_\omega^{static} &=& \int d{\bf r}\,\bigg[g_\omega\,\omega_\mu\,\bar\Psi\gamma^\mu\Psi\,-\,
\frac{1}{2}\,\bigg( m^{2}_{\omega}\,\omega^\mu\omega_\mu\,+\,\vec{\nabla}\omega^\mu\cdot \vec{\nabla}\omega_\mu
\,+\,(\vec{\nabla}\cdot\vec\omega)^2\bigg)\bigg]\nonumber\\
H_\rho^{static} &=& \int d{\bf r}\,\bigg[g_\rho\,\rho_{a\mu}\,\bar\Psi\gamma^\mu\tau_a\Psi\,+\,
\,g_\rho\frac{\kappa_\rho}{2\,M_N}\,\partial_j \rho_{a\mu}\,\bar\Psi\sigma^{\mu j}\tau_a\Psi
\nonumber\\
& & \,-\,\frac{1}{2}\,\bigg( m^{2}_{\rho}\,\rho_a^\mu\rho_{a\mu}\,+\,\vec{\nabla}\rho_a^\mu\cdot \vec{\nabla}\rho_{a\mu}
\,+\,(\vec{\nabla}\cdot\vec\rho_a)^2\bigg)\bigg]\nonumber\\
H_\delta^{static} &=& \int d{\bf r}\,\bigg[g_\delta\,\delta_a\,\bar\Psi\tau_a\Psi\,+\,
\frac{1}{2}\left((\vec{\nabla}\delta_a)^2\,+\,m^{2}_{\delta}\delta^{2}_{a}\right)\bigg]\nonumber\\
H_\pi^{static} &=& 	\int d{\bf r}\,\bigg[\frac{g_A}{2\,f_\pi}\,\vec{\nabla}\varphi_{a\pi}\cdot\bar\Psi\gamma^5\vec\gamma\tau_a\Psi
\,+\,\frac{1}{2}\left((\vec{\nabla}\varphi_{a\pi})^2\,+\,m^{2}_{\pi}\varphi^{2}_{a\pi}\right)\bigg].
\end{eqnarray}
This hamiltonian can be rewritten as~: 
\begin{equation}
H =	\int d{\bf r}\,\big(\,K\quad +\quad  H_{mesons}\big).\label{STATIC}
\end{equation}
The first term is the nucleonic piece including the Yukawa coupling of the nucleons to the meson fields~:
\begin{eqnarray}
K &=& \bar\Psi\bigg(-i\vec{\gamma}\cdot\vec{\nabla}\,+\,M_N(s)\,+\,g_\omega\,\omega_\mu\,\gamma^\mu\,
\,+\,g_\rho\,\rho_{a\mu}\,\gamma^\mu\tau_a\,+\,
\,g_\rho\frac{\kappa_\rho}{2\,M_N}\,\partial_j \rho_{a\mu}\,\sigma^{\mu j}\tau_a\nonumber\\
& &\,+\,g_\delta\,\delta_a\,\tau_a\,+\,\frac{g_A}{2\,f_\pi}\,\vec{\nabla}\varphi_{a\pi}\cdot\gamma^5\vec\gamma\tau_a\bigg)\Psi.\label{STATICK}
\end{eqnarray}
The second piece os of purely mesonic nature~: 
\begin{eqnarray}
H_{mesons}&=& \frac{1}{2}(\vec{\nabla}s)^2\,+\,V(s)
\,-\,
\frac{1}{2}\,\bigg( m^{2}_{\omega}\,\omega^\mu\omega_\mu\,+\,\vec{\nabla}\omega^\mu\cdot \vec{\nabla}\omega_\mu
\,+\,(\vec{\nabla}\cdot\vec\omega)^2\bigg)\nonumber\\
& & \,-\,\frac{1}{2}\,\bigg( m^{2}_{\rho}\,\rho_a^\mu\rho_{a\mu}\,+\,\vec{\nabla}\rho_a^\mu\cdot \vec{\nabla}\rho_{a\mu}
\,+\,(\vec{\nabla}\cdot\vec\rho_a)^2\bigg)\nonumber\\
& & \,+\,
\frac{1}{2}\left((\vec{\nabla}\delta_a)^2\,+\,m^{2}_{\delta}\delta^{2}_{a}\right)
\,+\,\frac{1}{2}\left((\vec{\nabla}\varphi_{a\pi})^2\,+\,m^{2}_{\pi}\varphi^{2}_{a\pi}\right).
\label{STATICM}
\end{eqnarray}

%%%%%%%%%%%%%%%%%%%%%%%%%%%%%%%%%%%%%%%%%%%%%%%%%%%%%%%%%%%%%%%%%%%%%%%%%%%%%
\subsection{Equation of motion for classical and fluctuating meson fields}
%%%%%%%%%%%%%%%%%%%%%%%%%%%%%%%%%%%%%%%%%%%%%%%%%%%%%%%%%%%%%%%%%%%%%%%%%%%%%%%%%%%%
In the static approximation the equation of 
motion for each meson field $\varphi_R$ can be written formally as $\left[H,\Pi_R\right]=0$, where $\Pi_R$ is the conjugate momentum 
of the field $\varphi_R$. This gives~:
\begin{eqnarray}
& & -\nabla^2 s\,+\,V'(s)=-\frac{\partial K}{\partial s}=
-\,\frac{\partial M_N}{\partial s}\,\bar\Psi\Psi\,\nonumber\\
& & -\nabla^2 \omega^\mu\,+\,m^{2}_{\omega}\,	\omega^\mu\,+\,\delta_{\mu i}\partial_i(\vec{\nabla}\cdot\vec\omega)
=\frac{\partial K}{\partial \omega_\mu}=g_\omega\,\bar\Psi\gamma^\mu\Psi\nonumber\\
& & -\nabla^2 \rho_a^\mu\,+\,m^{2}_{\rho}\,	\rho_a^\mu\,+\,\delta_{\mu i}\partial_i(\vec{\nabla}\cdot\vec\rho_a)
=\frac{\partial K}{\partial \rho_{a\mu}}=g_\rho\,\bar\Psi\gamma^\mu\tau_a\Psi\,-\,
g_\rho\frac{\kappa_\rho}{2\,M_N}\,\partial_j \left(\bar\Psi\sigma^{\mu j}\tau_a\Psi\right)
\nonumber\\
& & -\nabla^2 \delta_a\,+\,m^{2}_{\delta_a}\,	\delta
=\frac{\partial K}{\partial \delta_a}=g_\delta\,\bar\Psi\tau_a\Psi\nonumber\\
& & -\nabla^2 \varphi_{a\pi}\,+\,m^{2}_{\pi}\varphi_{a\pi}=\frac{\partial K}{\partial \varphi_{a\pi}}=
\frac{g_A}{2\,f_\pi}\,\vec{\nabla}\cdot\bar\Psi\gamma^5\vec\gamma\tau_a\Psi.
\end{eqnarray}
Following  the method used in ref. \cite{GMST06}  we now assume that it makes senses to decompose each meson field as~: 
\begin{equation}
\varphi_R=\bar\varphi_R\,+\,\Delta\varphi_R	
\end{equation}
where $\bar\varphi_R=\left\langle \varphi_R\right\rangle$ denotes the ground state expectation value 
of the meson field $\varphi_R$ and $\Delta\varphi_R$ corresponds to its fluctuation considered as a small quantity.

Since the case of the scalar field is the most delicate one we treat it below with some details.
The equation of motion for the $s$ field can be expanded in $\Delta s$ according to~:
\begin{equation}
-\nabla^2 (\bar s \,+\,\Delta s)\,+\,V'(\bar s)\,+\,\Delta s\,V''(\bar s)=-\frac{\partial K}{\partial s}	(\bar s)\,-\,
\Delta s\,\frac{\partial^2 K}{\partial s^2}	(\bar s).
\end{equation}
Explicitly we have~: 
\begin{eqnarray}
\frac{\partial K}{\partial s}	(\bar s) &\equiv &  \frac{\partial K}{\partial \bar s}=g^*_S\,\bar\Psi\Psi
\qquad\hbox{with}\qquad	g^*_S=\frac{\partial M_N(\bar s)}{\partial \bar s}=g_S\,+\,\kappa_{NS}\,\bar s\,+\,...\nonumber\\
\frac{\partial^2 K}{\partial s^2}	(\bar s) &\equiv &\frac{\partial^2 K}{\partial \bar s^2}=\tilde{\kappa}_{NS}\,\bar\Psi\Psi
\qquad\hbox{with}\qquad	\tilde{\kappa}_{NS}=\frac{\partial^2 M_N(\bar s^2)}{\partial \bar s}=\kappa_{NS}\,+\,...
\end{eqnarray}
We now develop the source term of the equation of motion according to~:
\begin{equation}
\frac{\partial K}{\partial \bar s}=\left\langle \frac{\partial K}{\partial \bar s}\right\rangle
+\Delta\left(\frac{\partial K}{\partial \bar s}\right)\equiv \left\langle \frac{\partial K}{\partial \bar s}\right\rangle\,+\,
\left(\frac{\partial K}{\partial \bar s}\,-\,\left\langle \frac{\partial K}{\partial \bar s}\right\rangle	\right)
\end{equation}
and we consider the fluctuating term $\Delta(\partial K/\partial \bar s)$ as small and of same order than $\Delta s$.
In the same spirit we replace the second derivative of $K$ by its expectation value~:
\begin{eqnarray}
\frac{\partial^2 K}{\partial \bar s^2}\approx\left\langle \frac{\partial^2 K}{\partial \bar s^2}\right\rangle=
\tilde{\kappa}_{NS}\,\left\langle \bar\Psi\Psi\right\rangle.
\end{eqnarray}
The equation of motion will be solved order by order~:
\begin{eqnarray}
& & -\nabla^2 \bar s \,+\,V'(\bar s)=-\left\langle \frac{\partial K}{\partial \bar s}\right\rangle=-g^*_S\,\left\langle \bar\Psi\Psi\right\rangle\nonumber\\
& & -\nabla^2 (\Delta s) \,+\,m^{*2}_{\sigma}\,\Delta s=-\left(\frac{\partial K}{\partial \bar s}\,-\,
\left\langle \frac{\partial K}{\partial \bar s}\right\rangle\right)=
-g^*_S\,\left(\bar\Psi\Psi\,-\,\left\langle \bar\Psi\Psi\right\rangle\right).\label{FLUCS}
\end{eqnarray}
In the equation for the fluctuating part it appears the effective scalar mass~:
\begin{equation}
m^{*2}_{\sigma}=V''(\bar s)\,+\,\tilde{\kappa}_{NS}\,\left\langle \bar\Psi\Psi\right\rangle	
\end{equation}
already introduced in our previous work \cite{CE05}. This is the physical in-medium scalar mass propagating the quantum fluctuations 
of the scalar field, {\it i.e.}, the quantum fluctuations of the chiral condensate in a non-trivial way.
In particular we will see below that it is the mass appearing in the Fock term of the scalar exchange at variance with $V'(\bar s)/\bar s$
appearing in the  Hartree scalar exchange. In that sense the treatment of the self-interacting scalar field deviates from the one of ref. 
\cite{BERNARD93}.
%%%%%%%%%%%%%%%%%%%%%%%%%%%%%%%%%%%%%%%%%%%%%%%%%%%%%%%%%%%%%%%%%%%%%%%%%%%%%%%%%%%%%%%%%%%%%%%%
\subsection{Kinetic, Hartree and exchange hamiltonians}
%%%%%%%%%%%%%%%%%%%%%%%%%%%%%%%%%%%%%%%%%%%%%%%%%%%%%%%%%%%%%%%
We now develop the hamiltonian (eq. \ref{STATIC}) to second order in the fluctuations, limiting ourselves to the nucleon field and the $s$ field, the generalization to the other mesons being straightforward.
\begin{eqnarray}
H_S &=& 	\int d{\bf r}\,\bigg[K(\bar s)\,+\,\Delta s\,\frac{\partial K}{\partial \bar s}\,+\,\frac{1}{2}\,
(\Delta s)^2\,\frac{\partial^2 K}{\partial \bar s^2}\,+
\,\frac{1}{2}\,\left(-\bar s\nabla^2\bar s\,-\,2\,\Delta s\nabla^2\bar s\,+\,(\vec{\nabla}(\Delta s))^2\right)\nonumber\\
& &\,+\,V(\bar s)\,+\,\Delta s\,V'(\bar s)\,+\,\frac{1}{2}\,(\Delta s)^2\,V''(\bar s)\bigg].
\end{eqnarray}
Using the classical equation of motion and the one for the fluctuating field and replacing again the second derivative of $K$ by its expectation value, we obtain~: 
\begin{eqnarray}
H_S &=& 	\int d{\bf r}\,\bigg[\bar\Psi\left(-i\vec{\gamma}\cdot\vec{\nabla}\,+\,M_N(\bar s)\right)\Psi\,
+\,\frac{1}{2}\,\big(\vec{\nabla}(\bar s)\big)^2\,+\,V(\bar s)
\,+\,\frac{1}{2}\,g^*_S\,\left(\bar\Psi\Psi\,-\,\left\langle \bar\Psi\Psi\right\rangle\right)\,\Delta s\bigg]\nonumber\\
& & \label{HS}
\end{eqnarray}
The approach can be extended to the other mesons. Only the delta field and the time component of the rho and omega mesons
have a non zero expectation value, solution of the classical equations~:
\begin{eqnarray}
& & -\nabla^2 \bar\omega^0\,+\,m^{2}_{\omega}\,	\bar\omega^0=g_\omega\,\left\langle \Psi^\dagger\Psi\right\rangle\nonumber\\
& & -\nabla^2 \bar\rho_a^0\,+\,m^{2}_{\rho}\,	\bar\rho_a^0\,=g_\rho\,\left\langle \Psi^\dagger\tau_a\Psi\right\rangle
\,-\,g_\rho\frac{\kappa_\rho}{2\,M_N}\,\partial_j \left\langle \bar\Psi\sigma^{0 j}\tau_a\Psi\right\rangle\nonumber\\
& & -\nabla^2 \bar\delta_a\,+\,m^{2}_{\delta_a}\,	\bar\delta
=g_\delta\,\left\langle \bar\Psi\tau_a\Psi\right\rangle	.
\end{eqnarray}
The fluctuating fields are solutions of~:
\begin{eqnarray}
& & -\nabla^2 (\Delta\omega^\mu)\,+\,m^{2}_{\omega}\,	\Delta\omega^\mu=P^{\mu}_{\;\nu}\,
g_\omega\,\left(\bar\Psi\gamma^\nu\Psi\,-\,\left\langle  \bar\Psi\gamma^\nu\Psi\right\rangle\right)\nonumber\\
& & -\nabla^2 (\Delta\rho_a^\mu)\,+\,m^{2}_{\rho}\,	\Delta\rho^\mu=	P^{\mu}_{\;\nu}\,
g_\rho\,\bigg(\bar\Psi\gamma^\nu\tau_a\Psi\,-\,\left\langle \bar\Psi\gamma^\nu\tau_a\Psi\right\rangle\nonumber\\
& & 
\qquad\qquad -\,\frac{\kappa_\rho}{2\,M_N}\,\partial_j \left(\bar\Psi\sigma^{\nu j}\tau_a\Psi\right)
\,+\,\frac{\kappa_\rho}{2\,M_N}\,\partial_j \left\langle\bar\Psi\sigma^{\nu j}\tau_a\Psi\right\rangle
\bigg)\nonumber\\
& & -\nabla^2 (\Delta\delta_a)\,+\,m^{2}_{\delta}\,	\Delta\delta_a
=g_\delta\,\left(\bar\Psi\tau_a\Psi\,-\,\left\langle \bar\Psi\tau_a\Psi\right\rangle\right)\nonumber\\
& & -\nabla^2 (\Delta\varphi_{a\pi})\,+\,m^{2}_{\pi}\,\Delta\varphi_{a\pi}
=\frac{g_A}{2\,f_\pi}\,\vec{\nabla}\cdot\bar\Psi\gamma^5\vec\gamma\tau_a\Psi
\end{eqnarray}
with~:
\begin{equation}
P^{0}_{\;0}=1,\quad	P^{0}_{\;i}=0,\quad P^{i}_{\;0}=0,\quad P^{i}_{\;j}\equiv P^{i}_{\;j}(x)=\delta_{ij}\,-\,\frac{\partial_i\partial_j}{m^{2}_{\rho}}.
\end{equation}
Generalizing the result of eq. (\ref{HS}) to all the mesons, the full hamiltonian, in the static approximation,  can be written in a form reminiscent of the density functional theory~:
\begin{equation}
H=H_{kin+Hartree}\,+\,H_{xc}.	
\end{equation}
The first term is a one-body operator containing  the kinetic energy hamiltonian of the nucleons and 
the other pieces of the hamiltonian contributing to the Hartree energy. Its explicit form is~:
\begin{eqnarray}
H_{kin+Hartree}&=& \int d{\bf r}\,\bigg[\bar\Psi\bigg(-i\vec{\gamma}\cdot\vec{\nabla}\,+\,M_N(\bar s)\,+\,g_\omega\,
\bar\omega^0\,\gamma_0\,+\,g_\rho\,\bar\rho_3^0\,\gamma_0\tau_3\nonumber\\
& &\,+\,g_\rho\frac{\kappa_\rho}{2\,M_N}\,\partial_j \bar\rho^0_3\,\sigma^{0 j}\tau_3\,+\,g_\delta\,
\bar\delta_3\,\tau_3\bigg)\Psi\nonumber\\ 
& & \,+	\,V(\bar s)\,+\,\frac{1}{2}\,
\left(\vec{\nabla}\bar s\right)^2\,-\,\frac{1}{2}\,m^2_\omega\, (\bar\omega^0)^2\,-\,
\frac{1}{2}\left(\vec{\nabla}\bar\omega^0\right)^2\nonumber\\
& &\,-\,\frac{1}{2}\,m^2_\rho\,(\bar\rho_3^0)^2
\,-\,\frac{1}{2}\left(\vec{\nabla}\bar\rho^0_3\right)^2
\,+\,\frac{1}{2}\,m^2_\delta\,\bar\delta_3^2
\,+\,\frac{1}{2}\left(\vec{\nabla}\bar\delta_3\right)^2\bigg].
\nonumber\\
\end{eqnarray}
Considering only this part of the  hamiltonian we come to the conclusion that 
symmetric and asymmetric nuclear matter are  seen as an assembly of nucleons, {\it i.e.,} of Y shaped color strings 
with massive constituent quarks at the end getting their mass from the chiral condensate.
This nucleons move in self-consistent scalar ($\bar s, \bar\delta_3$) and vector background fields 
($\bar\omega^0, \bar\rho^0_3$). The scalar field which we associate with 
the radial mode of the condensate modifies the nucleon mass according to 
$M_N(\bar s)=M_N\,+\,g_S\,\bar s\, +\,\kappa_{NS}\,\bar s^2/2\,+\,..$ .	
$\,g_S\,\bar s$ describes the in-medium  dropping of the nucleon mass from its coupling to the in-medium modified chiral condensate but there is another term, $\kappa_{NS}\,\bar s^2$, which corresponds to the  response of the nucleon to the background scalar field. The scalar response $\kappa_{NS}$ depends on the structure of the nucleon and takes into account  the modification of the quark wave functions inside the nucleon and obviously depends on the confinement mechanism. The nucleon is also submitted to an isovector field, ($\bar\omega^0$) and in asymmetric nuclear matter
to an isovector vector field ($\bar\rho^0_3$) or even an isovector scalar field ($\bar\delta_3$). 

The second piece of the hamiltonian, $H_{xc}$ incorporates the exchange term mediated by the propagation of the fluctuations of the meson fields.
In particular the scalar fluctuation, {\it i.e.}, the fluctuation of the chiral condensate propagates, as already stated, with an in-medium modified sigma mass $m^{*2}_{\sigma}=V''(\bar s)\,+\,\tilde{\kappa}_{NS}\,\rho_S$. Its explicit form is~:
\begin{eqnarray}
H_{xc} &=&	\int d{\bf r}\,\frac{1}{2}\,\bigg[g^*_S\,\,\Delta s\,\Delta\left(\bar\Psi\Psi\right)
\,+\,g_\omega\,\,\Delta \omega_\mu\left(\bar\Psi\gamma^\mu\Psi\right)\nonumber\\
& &\,+\,g_\rho\,\Delta\rho^{a}_\mu\,\bigg(\Delta(\bar\Psi\gamma^\mu\tau_a\Psi)\,-\,
\frac{\kappa_\rho}{2\,M_N}\,\partial_j \left[\Delta\left(\bar\Psi\sigma^{\mu j}\tau_a\Psi\right)\right]\bigg)\nonumber\\
& &\,+\,g_\delta\,\Delta\delta_a\,\Delta\left(\bar\Psi\tau_a\Psi\right)
\,+\,\frac{g_A}{2\,f_\pi}\,\Delta\varphi_{a\pi}\,\vec{\nabla}\cdot\bar\Psi\gamma^5\vec\gamma\tau_a\Psi\bigg]
\end{eqnarray}
where  we have used systematically the notation~:
$$\Delta(\bar\Psi\Gamma\Psi)=\bar\Psi\Gamma\Psi\,-\,\left\langle \bar\Psi\Gamma\Psi\right\rangle.$$
We introduce the (static) propagators for the fluctuating fields~~:
\begin{eqnarray}
& & \left(-\nabla^{2}_{\bf r}\,+\,m^{*2}_{\sigma}({\bf r})\right)\,D_\sigma({\bf r}\,-\,{\bf r}')=\delta^{(3)}({\bf r}\,-\,{\bf r}')\nonumber\\
& & \left(-\nabla^{2}_{\bf r}\,+\,m^{2}_{\alpha}\right)\,D_\alpha({\bf r}\,-\,{\bf r}')=	\delta^{(3)}({\bf r}\,-\,{\bf r}'),
\qquad \alpha=\omega, \rho, \delta, \pi\label{PROPS}
\end{eqnarray}
and solve formally for the fluctuating fields. $H_{xc}$ can be written as~:
\begin{eqnarray}
H_{xc}&=& \frac{1}{2}\int d{\bf r}\,	d{\bf r}' \bigg[
-\,g^*_S({\bf r})\,g^*_S({\bf r}')\,\Delta\left(\bar\Psi\Psi\right)({\bf r})\,D_\sigma({\bf r}-{\bf r}')\,
\Delta\left(\bar\Psi\Psi\right)({\bf r}')\nonumber\\
& &\,+\,g^{2}_\omega\,\Delta\left(\bar\Psi\gamma^\mu\Psi\right)({\bf r})
\, D_{\omega\mu\nu}({\bf r}-{\bf r}')\,
\Delta\left(\bar\Psi\gamma^\nu\Psi\right)({\bf r}')\nonumber\\
& &\,+\,g^{2}_\rho\,\Delta\left(\bar\Psi\gamma^\mu\tau_a\Psi\right)({\bf r})
\, D_{\rho\mu\nu}({\bf r}-{\bf r}')\,
\Delta\left(\bar\Psi\gamma^\nu\tau_a\Psi\right)({\bf r}')\nonumber\\
& & \,+\,2\,g^{2}_\rho\frac{\kappa_\rho}{2\,M_N}\, \Delta\left(\bar\Psi\sigma^{\mu j}\tau_a\Psi\right)({\bf r})
\, \partial_j D_{\rho\mu\nu}({\bf r}-{\bf r}')\,
\Delta\left(\bar\Psi\gamma^\nu\tau_a\Psi\right)({\bf r}')\nonumber\\
& & \,+\,g^{2}_\rho\left(\frac{\kappa_\rho}{2\,M_N}\right)^2\, \Delta\left(\bar\Psi\sigma^{\mu i}\tau_a\Psi\right)({\bf r})
 \partial_i\partial'_j D_{\rho\mu\nu}({\bf r}-{\bf r}')\,
\Delta\left(\bar\Psi\sigma^{\nu j}\tau_a\Psi\right)({\bf r}')\nonumber\\
& &\,-\,g^2_\delta\,
\Delta\left(\bar\Psi\tau_a\Psi\right)({\bf r})
\,D_\delta({\bf r}-{\bf r}')\,
\Delta\left(\bar\Psi\tau_a\Psi\right)({\bf r}')\nonumber\\
& & \,+\,\left(\frac{g_A}{2\,f_\pi}\right)^2\,\left(\bar\Psi\gamma^5\gamma^i\tau_a\Psi\right)({\bf r})
\,\partial_i\partial'_j D_\pi({\bf r}-{\bf r}')\,
\left(\bar\Psi\gamma^5\gamma^j\tau_a\Psi\right)({\bf r}')\bigg]
\end{eqnarray}
 where we have introduced the tensor propagator $D_{\omega\mu\nu}({\bf r}-{\bf r}')$ whose non vanishing components are~:
 $D_{\omega 00}({\bf r}-{\bf r}')=D_{\omega}({\bf r}-{\bf r}')$ and
 $D_{\omega ij}({\bf r}-{\bf r}')=(\delta_{ij}\,-\,\partial_{i}\partial_{j}/m^2_\omega)\,D_{\omega}({\bf r}-{\bf r}')$ and a similar 
 one for the rho meson.
%%%%%%%%%%%%%%%%%%%%%%%%%%%%%%%%%%%%%%%%%%%%%%%%%%%%%%%%%%%%%%%%%%%%%%%%%%%
\section{Hartree-Fock approach}\label{HAR}
%%%%%%%%%%%%%%%%%%%%%%%%%%%%%%%%%%%%%%%%%%%%%%%%%%%%%%%%%%%%%%%%%%%%%%%%%
\subsection{Hartree-Fock energy}
In the Hartree-Fock approximation the ground state is represented by a Slater determinant made of single particle states with Dirac wave functions 
$\varphi_a^{N}(\vec{r})\,\chi_N$, with $N=p,n$ for protons and neutrons. The various densities which appear in the sources of the classical equations of motions are~:
\begin{eqnarray}
\left\langle \bar{\Psi}\Psi\right\rangle &=& \sum_{a<F}\,\bar{\varphi}_a^{p}\varphi^{p}_{a}\,+\,
\bar{\varphi}_a^{n}\varphi^{n}_{a}\equiv\rho_{Sp}\,+\,\rho_{Sn}\nonumber\\
\left\langle \Psi^\dagger\Psi\right\rangle &=& \sum_{a<F}\,\varphi_a^{p\dagger}\varphi^{p}_{a}\,+\,
\varphi_a^{n\dagger}\varphi^{n}_{a}\equiv\rho_{p}\,+\,\rho_{n}\nonumber\\
\left\langle \bar{\Psi}\tau_3\Psi\right\rangle &=& \sum_{a<F}\,\bar{\varphi}_a^{p}\varphi^{p}_{a}\,-\,
\bar{\varphi}_a^{n}\varphi^{n}_{a}\equiv\rho_{Sp}\,-\,\rho_{Sn}\nonumber\\
\left\langle \Psi^\dagger\tau_3\Psi\right\rangle &=& \sum_{a<F}\,\varphi_a^{p\dagger}\varphi^{p}_{a}\,-\,
\varphi_a^{n\dagger}\varphi^{n}_{a}\equiv\rho_{p}\,-\,\rho_{n}\nonumber\\
\left\langle \bar{\Psi}\sigma^{0 j}\tau_3\Psi\right\rangle &=& \sum_{a<F}\,\bar{\varphi}_a^{p}\sigma^{0 j}\varphi^{p}_{a}\,-\,
\bar{\varphi}_a^{n}\sigma^{0 j}\varphi^{n}_{a}
\end{eqnarray}
The kinetic plus Hartree piece of the total energy is~:
 \begin{eqnarray}
E_{kin+H}&=& \int d{\bf r}\,\bigg[\sum_{a<F}\,\bar\varphi_a^{p}(-i\vec{\gamma}\cdot\vec{\nabla}\,+\,M_N(\bar s)\,+\,g_\omega\,
\bar\omega^0\,\gamma_0\,+\,g_\rho\,\bar\rho_3^0\,\gamma_0\nonumber\\
& &\,+\,i\,g_\rho\frac{\kappa_\rho}{2\,M_N}\,\vec{\nabla}\bar\rho^0_3\cdot \vec{\Sigma}\,\gamma_5\,+\,g_\delta\,
\bar\delta_3\,\bigg)\varphi_a^p\bigg]\nonumber\\ 
& &\,+\int d{\bf r}\,\bigg[\sum_{a<F}\,\bar\varphi_a^{n}(-i\vec{\gamma}\cdot\vec{\nabla}\,+\,M_N(\bar s)\,+\,g_\omega\,
\bar\omega^0\,\gamma_0\,-\,g_\rho\,\bar\rho_3^0\,\gamma_0\nonumber\\
& &\,-\,i\,g_\rho\frac{\kappa_\rho}{2\,M_N}\,\vec{\nabla}\bar\rho^0_3\cdot \vec{\Sigma}\,\gamma_5\,-\,g_\delta\,
\bar\delta_3\,\bigg)\varphi_a^n\bigg]\nonumber\\ 
& & \,+	\,\int d{\bf r}\,\bigg[V(\bar s)\,+\,\frac{1}{2}\,
\left(\vec{\nabla}\bar s\right)^2\,-\,\frac{1}{2}\,m^2_\omega\, (\bar\omega^0)^2\,-\,
\frac{1}{2}\left(\vec{\nabla}\bar\omega^0\right)^2\nonumber\\
& &\,-\,\frac{1}{2}\,m^2_\rho\,(\bar\rho_3^0)^2
\,-\,\frac{1}{2}\left(\vec{\nabla}\bar\rho^0_3\right)^2
\,+\,\frac{1}{2}\,m^2_\delta\,\bar\delta_3^2
\,+\,\frac{1}{2}\left(\vec{\nabla}\bar\delta_3\right)^2\bigg].
\label{HARTRENERGY}
\end{eqnarray}
The Fock term contribution  to the energy comes entirely from $H_{xc}$. It can be split as~:
\begin{equation} 
E_{Fock}=E_{Fock}^{(s)}\,+\,E_{Fock}^{(\omega)}\,+\,E_{Fock}^{(\rho)}\,+\,E_{Fock}^{(\delta)}\,+\,E_{Fock}^{(\pi)}\,.
\label{FOCKENERG}
\end{equation}
We give here the expression for the scalar field piece
\begin{eqnarray}
E_{Fock}^{(s)}&=&\frac{1}{2\,}\int\,d{\bf r}\,d{\bf r}'\,Tr\big(S_p({\bf r}'-{\bf r})\,  S_p({\bf r}-{\bf r}')\,+\,
S_n({\bf r}'-{\bf r})\,  S_n({\bf r}-{\bf r}')\big)\nonumber\\
& &\qquad\qquad\qquad g^*_S({\bf r})\,g^*_S({\bf r}')\,D_\sigma({\bf r}-{\bf r}'),\label{FOCKS}
\end{eqnarray}
the other contribution being given in appendix A.
Here $S_p$ and $S_n$  are matrices in Dirac space~:
$$\big(S_p({\bf r}-{\bf r}')\big)_{\alpha\beta}=\sum_{a<F}\,\left(\varphi^p_a\right)_{\alpha}({\bf r})\,
\left(\bar\varphi^p_a\right)_{\beta}({\bf r}')\qquad
\big(S_n({\bf r}-{\bf r}')\big)_{\alpha\beta}=\sum_{a<F}\,\left(\varphi^n_a\right)_{\alpha}({\bf r})\,
\left(\bar\varphi^n_a\right)_{\beta}({\bf r'}).$$
%%%%%%%%%%%%%%%%%%%%%%%%%%%%%%%%%%%%%%%%%%%%%%%%%%%%%%%%%%%%%%%%%%%%%%%%%%%%%%%%%%%%%%%%%%%%%%%%%%%%%%%%%%%%%%%%%%%
\subsection{Hartree-Fock equations}
%%%%%%%%%%%%%%%%%%%%%%%%%%%%%%%%%%%%%%%%%%%%%%%%%%%%%%%%%%%%%%%%%%%%%%%%%%%%%%%%%%%%%%%%%%%%%%%%%%%%%%% 
The single particle orbitals are obtained by minimizing the HF energy with respect to the $\bar\varphi^{p,n}_a({\bf x})$, with the constraint that the single particle wave functions are normalized, {\it i.e.},
\begin{equation}
\frac{\delta \bigg(E\,-\,\sum_{N,a}\varepsilon_a^N\, \int d{\bf r}\varphi^{N\dagger}_a({\bf r})\varphi^{N}_a({\bf r})\bigg)}{\delta \bar\varphi^{p,n}_a({\bf x}) }=0
\end{equation}
and the Lagrange parameters $\varepsilon_a^N$ have to be identified with the single-particle energies.
According to the results of the previous section the structure of this HF energy is~: 
\begin{eqnarray}
E &=& \sum_{a,N}\int d{\bf r}\,\bigg[\,\bar\varphi_a^{N}\bigg(-i\vec{\gamma}\cdot\vec{\nabla}\,+\,M_N(\bar s)\,+\,\sum_R g^{(N)}_R\,\bar\varphi_R\,
\Gamma_R\,+\,i\,h_R^{(N)}\,\vec{\nabla}\bar\varphi_R\,\cdot\vec{\Theta}_R\bigg)\varphi_a^{N}\bigg]\nonumber\\
& &\quad\quad +\,E_M(\bar\varphi_R, \vec{\nabla}\bar\varphi_R)\nonumber\\
& &\, +\sum_{R,R';N,N'}\frac{g_{R R'}^{(N N')}}{2}\,\int\,d{\bf r}\,d{\bf r}'\,
Tr\big(S_N({\bf r}'-{\bf r})  
\,\Gamma_R\,S_{N'}({\bf r}-{\bf r}')\,\Gamma_{R'}\bigg)\,D_{R R'}({\bf r}-{\bf r}')
\end{eqnarray}
The first two lines correspond to the kinetic plus Hartree energy (eq. \ref{HARTRENERGY}), the second line  being  the purely mesonic piece;  the last line  represents  the Fock (exchange) term (eq. \ref{FOCKENERG}). The various coupling constants, $g^{(N)}_R, h_R^{(N)}, g_{R R'}^{(N N')}$, Dirac operators, $\Gamma_R, \vec{\Theta}_R$, and propagators,
$D_{R R'}({\bf r}-{\bf r}')$, are obtained in an obvious way by direct comparison with the explicit expressions of the energy given above 
(eqs. \ref{HARTRENERGY}, \ref{FOCKENERG}, \ref{FOCKS} and Appendix A).
For the the particular case of the scalar channel the coupling $g_{R R'}^{(N N')}$ has to be inserted inside  the integral and replaced by 
$g^*_S({\bf r})\,g^*_S({\bf r}')$. In the minimization procedure the derivative of
the expectation values of the mesonic fields $\bar\varphi_R$ is not taken since it is already accounted for by the equations of motion. 
The last line of the HF energy actually generates two kinds of contributions in the HF equations; the first one comes from the functional derivative of the nucleon propagators, $S_{N}({\bf r}-{\bf r}')$  and the second one 
comes from the derivative of $g^*_S\,g^*_S\,D_\sigma$ in the scalar channel (eq. \ref{FOCKS}). The later is actually a rearrangement term associated with the many body forces originating from the in-medium dependence of the scalar meson mass and of the scalar coupling constant.
\smallskip\noindent
Ignoring for the moment the rearrangement terms, the Hartree-Fock equations writes~:
\begin{equation}
\int d{\bf r}'\,\left\langle{\bf r}\left|h^N_{(ord)}\right|{\bf r}'\right\rangle\,\varphi_a^{N}({\bf r}')=\varepsilon_a^N\,\varphi_a^{N}({\bf r})
\end{equation}
where the single particle hamiltonian, $h^N$, which is represented by a matrix in Dirac space, has a local piece (the direct Hartree piece) and a non local piece (the Fock exchange term). With the previous schematic notations it is such that ~: 
\begin{eqnarray}
& &\left\langle {\bf r}\left|\gamma^0\,h^N_{(ord)}\right|{\bf r}'\right\rangle\nonumber\\
& &\quad =\bigg(-i\vec{\gamma}\cdot\vec{\nabla}\,+\,M_N(\bar s)\,+\,\sum_R g^{(N)}_R\,\bar\varphi_R({\bf r})\,
\Gamma_R\,+\,i\,h_R^{(N)}\,\vec{\nabla}\bar\varphi_R({\bf r})\,\cdot\vec{\Theta}_R\bigg)\delta^{(3)}({\bf r}-{\bf r}')\nonumber\\
& &\quad +\sum_{R,R';N'}\frac{g_{R R'}^{(N N')}}{2}\,D_{R R'}({\bf r}-{\bf r}')\,\bigg(\Gamma_R\,S_{N'}({\bf r}-{\bf r}')\,\Gamma_{R'}\bigg)\nonumber\\
& &\quad +\sum_{R,R';N'}\frac{g_{R R'}^{(N N')}}{2}\,D_{R R'}({\bf r}'-{\bf r})\bigg(\Gamma_{R'}\,S_{N'}({\bf r}-{\bf r}')\,\Gamma_{R}\bigg)
\label{ORDSE}.
\end{eqnarray}
\begin{figure}
\noindent
 \begin{minipage}[b]{.450\linewidth}
    \centering\epsfig{figure=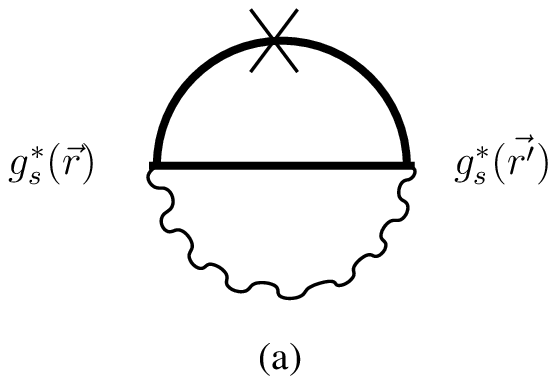,width=\linewidth}
  \end{minipage}\hfill
  \begin{minipage}[b]{.45\linewidth}. 
  \centering\epsfig{figure=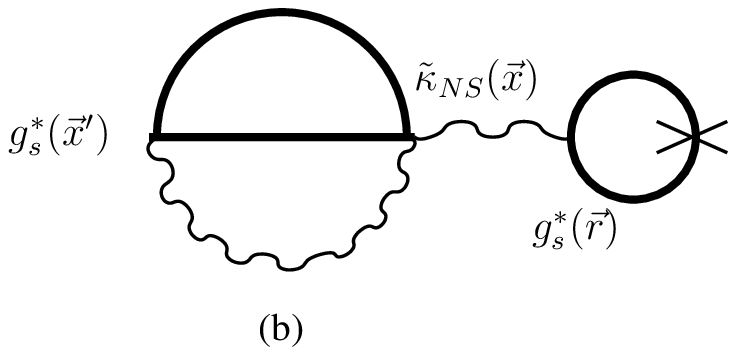,width=\linewidth}
  \end{minipage}
 \begin{minipage}[b]{.30\linewidth}
    \centering\epsfig{figure=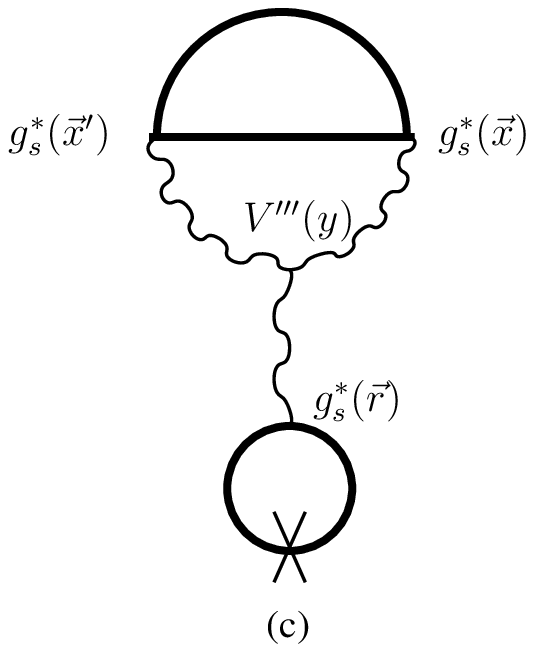,width=\linewidth}
  \end{minipage}\hfill
  \begin{minipage}[b]{.30\linewidth}. 
  \centering\epsfig{figure=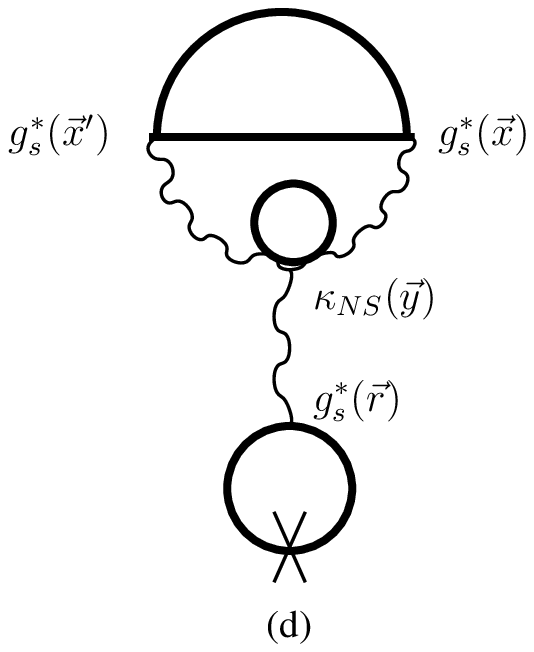,width=\linewidth}
  \end{minipage}
  \begin{minipage}[b]{.30\linewidth}. 
  \centering\epsfig{figure=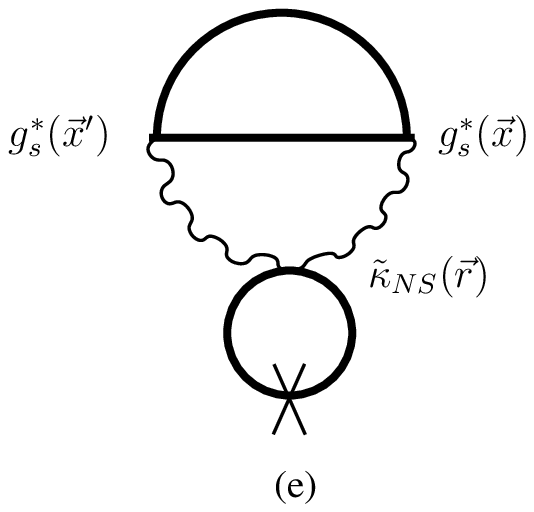,width=\linewidth}
  \end{minipage}
\caption{ Rearrangement terms; the crosses  indicate the opening of a line to generate the various contributions to the single-particle hamiltonian.  Detailed  explanations are given in the text. } 
\end{figure}
\smallskip\noindent
Let us now come to the rearrangement term originating from the density dependence of the scalar mass and coupling constant. Its contribution 
to the single particle hamiltonian is defined by~: 
\begin{eqnarray}
\left\langle {\bf r}\left|\gamma^0\,h^N_{(rg)} \right|\varphi_a^{N}\right\rangle &=&
\frac{1}{2}\int d{\bf x}\,d{\bf x}'\,\sum_M Tr\bigg(S_M({\bf x}\,-\,{\bf x}')\,S_M({\bf x}'\,-\,{\bf x})\bigg)\nonumber\\
& &\qquad\frac{\delta}{\delta\bar\varphi^{N}_{a}({\bf r})}\bigg(g^{*}_{S}(\bf x})\,g^{*}_{S}({\bf x}')\,
D_\sigma({\bf x}\,-\,{{\bf x}')\bigg).\label{REAR}
\end{eqnarray}
To get the functional derivative of the scalar coupling constant we start from
\begin{equation}
\frac{\delta g^{*}_{S}({\bf x})}{\delta\bar\varphi^{N}_{a}({\bf r})}=\tilde{\kappa}_{NS}\,({\bf x})\,
\frac{\delta\bar s({\bf x})}{\delta\bar\varphi^{N}_{a}({\bf r})}
\end{equation}
and take the functional derivative of the equation determining $\bar s$ (first equation of eq. (\ref{FLUCS}))~:
\begin{equation}
\bigg(-\nabla^2_{\bf x}\,+\,V''(\bar s({\bf x}))\,+\,\tilde{\kappa}_{NS}({\bf x})\,\rho_S({\bf x})\,\bigg)
\frac{\delta\bar s({\bf x})}{\delta\bar\varphi^{N}_{a}({\bf r})}=-\,g^{*}_{S}({\bf x})\,\varphi^{N}_{a}({\bf r})\,\delta^{(3)}({\bf x}\,-\,{\bf r})
\end{equation}
Notice that the full in-medium sigma mass and the full inverse sigma propagator $D^{-1}_\sigma$ appear in the left-hand side of the above equation. The solution of this equation is thus~:
\begin{equation}
\frac{\delta\bar s({\bf x})}{\delta\bar\varphi^{N}_{a}({\bf r})}=-\int d{\bf x}'\,D_\sigma({\bf x}\,-\,{\bf x}')
g^{*}_{S}({\bf x}')\,\varphi^{N}_{a}({\bf r})\,\delta({\bf x}'\,-\,{\bf r})=
-\,D_\sigma({\bf x}\,-\,{\bf r})\,g^{*}_{S}({\bf r})\,\varphi^{N}_{a}({\bf r})
\end{equation}
It follows that the derivative of the scalar coupling constant has the explicit form~:
\begin{equation}
\frac{\delta g^{*}_{S}({\bf x})}{\delta\bar\varphi^{N}_{a}({\bf r})}	=-\tilde{\kappa}_{NS}\,({\bf x})\,	
D_\sigma({\bf x}\,-\,{\bf r})\,g^{*}_{S}({\bf r})\,\varphi^{N}_{a}({\bf r}).
\end{equation}
Similarly the functional derivative of the sigma propagator can be obtained by taking the functional derivative of eq. (\ref{PROPS}).
One obtains~: 
\begin{equation}
\bigg(-\nabla^{2}_{\bf x}\,+\,m^{*2}_{\sigma}({\bf x})\bigg)
\frac{\delta D_{\sigma}({\bf x}\,-\,{\bf x}')}{\delta\bar\varphi^{N}_{a}({\bf r})}=-\,
\frac{\delta m^{*}_{\sigma}({\bf x})}{\delta\bar\varphi^{N}_{a}({\bf r})}\,D_{\sigma}({\bf x}\,-\,{\bf x}').
\end{equation}
The functional derivative of $m^{*2}_{\sigma}({\bf x}))=V''(\bar s({\bf x}))\,+\,\tilde{\kappa}_{NS}({\bf x})\,\rho_S({\bf x})$ is given by~:
\begin{eqnarray}
\frac{\delta m^{*}_{\sigma}({\bf x})}{\delta\bar\varphi^{N}_{a}({\bf r})}&=&
\left(V'''(\bar s)\,+\,\frac{\partial\tilde{\kappa}_{NS}}{\partial \bar s}\rho_S\right)({\bf x})\,
\frac{\delta\bar s({\bf x})}{\delta\bar\varphi^{N}_{a}({\bf r})}\,+\,
\tilde{\kappa}_{NS}({\bf x})\,\varphi^{N}_{a}({\bf r})\,\,\delta^{(3)}({\bf x}\,-\,{\bf r})\nonumber\\
&=& \left(V'''(\bar s)\,+\,\frac{\partial\tilde{\kappa}_{NS}}{\partial \bar s}\rho_S\right)({\bf x})\,
\bigg(-\,D_\sigma({\bf x}\,-\,{\bf r})\,g^{*}_{S}({\bf r})\,\varphi^{N}_{a}({\bf r})\bigg)\nonumber\\
& &\,+\,\tilde{\kappa}_{NS}({\bf x})\,\varphi^{N}_{a}({\bf r})\,\,\delta^{(3)}({\bf x}\,-\,{\bf r}).
\end{eqnarray}
The explicit form of the functional derivative of the in-medium sigma propagator follows~:
\begin{eqnarray}
& &\frac{\delta D_{\sigma}({\bf x}\,-\,{\bf x}')}{\delta\bar\varphi^{N}_{a}({\bf r})}=
-\int d{\bf y}\,D_{\sigma}({\bf x}\,-\,{\bf y})\,\frac{\delta m^{*}_{\sigma}({\bf y})}{\delta\bar\varphi^{N}_{a}({\bf r})}
D_{\sigma}({\bf y}\,-\,{\bf x}')\nonumber\\
& &=\int d{\bf y}\,D_{\sigma}({\bf x}\,-\,{\bf y})\,
\left(V'''(\bar s)\,+\,\frac{\partial\tilde{\kappa}_{NS}}{\partial \bar s}\rho_S\right)({\bf y})\,D_\sigma({\bf y}\,-\,{\bf x}')\,
D_\sigma({\bf y}\,-\,{\bf r})\,g^{*}_{S}({\bf r})\,\varphi^{N}_{a}({\bf r})\nonumber\\
& & \qquad -\, D_\sigma({\bf x}\,-\,{\bf r})\,D_\sigma({\bf r}\,-\,{\bf x}')\,\tilde{\kappa}_{NS}{\bf r})\,\varphi^{N}_{a}({\bf r}).
\end{eqnarray}
The rearrangement single particle hamiltonian (eq.\ref{REAR}) can be decomposed in two terms. The first one comes from the functional derivative of $g^{*}_{S}({\bf x})$~: 
\begin{eqnarray}
\left\langle {\bf r}\left|\gamma^0\,h^N_{(rg\,1)} \right|\varphi_a^{N}\right\rangle &=&
-\,\frac{1}{2}\int d{\bf x}\,d{\bf x}'\,Tr_M \bigg(S_M({\bf x}\,-\,{\bf x}')\,S_M({\bf x}'\,-\,{\bf x})\bigg)\,
D_\sigma({\bf x}\,-\,{\bf x}')\nonumber\\
& &\qquad\bigg({\tilde\kappa}_{NS}({\bf x})\,g^{*}_{S}({\bf x}')\,
D_\sigma({\bf x}\,-\,{\bf r})\,g^{*}_{S}({\bf r})\,\varphi^{N}_{a}({\bf r})\nonumber\\
& &\qquad \,+\,({\tilde\kappa}_{NS}({\bf x}')\,g^{*}_{S}({\bf x})\,
D_\sigma({\bf x}'\,-\,{\bf r})\,g^{*}_{S}({\bf r})\,\varphi^{N}_{a}({\bf r})\bigg).
\label{REAR1}
\end{eqnarray}
The second term originates from the derivative of the sigma propagator~:
\begin{eqnarray}
& &\left\langle {\bf r}\left|\gamma^0\,h^N_{(rg\,2)} \right|\varphi_a^{N}\right\rangle =
\int d{\bf x}\,d{\bf x}'\,Tr_M \bigg(S_M({\bf x}\,-\,{\bf x}')\,S_M({\bf x}'\,-\,{\bf x})\bigg)\,
\,g^{*}_{S}({\bf x})\,\,g^{*}_{S}({\bf x}')\nonumber\\
& &\qquad\bigg[\int d{\bf y}\,D_{\sigma}({\bf x}\,-\,{\bf y})\,
\left(V'''(\bar s)\,+\,\frac{\partial\tilde{\kappa}_{NS}}{\partial \bar s}\rho_S\right)({\bf y})\,D_\sigma({\bf y}\,-\,{\bf x}')\,
D_\sigma({\bf y}\,-\,{\bf r})\,g^{*}_{S}({\bf r})\nonumber\\
& &\qquad -\,D_\sigma({\bf x}\,-\,{\bf r})\,D_\sigma({\bf r}\,-\,{\bf x}')\,\tilde{\kappa}_{NS}{\bf r})\bigg]\,\varphi^{N}_{a}({\bf r}).
\end{eqnarray}
Notice that these rearrangement terms are of local nature, hence the Hartree-Fock equations finally write~:
\begin{equation}
\int d{\bf r}'\,\left\langle{\bf r}\left|h^N_{(ord)}\right|{\bf r}'\right\rangle\,\varphi_a^{N}({\bf r}')\,+\,
\left\langle {\bf r}\left|\gamma^0\,h^N_{(rg\,1)} \right|\varphi_a^{N}\right\rangle \,+\,
\left\langle {\bf r}\left|\gamma^0\,h^N_{(rg\,2)} \right|\varphi_a^{N}\right\rangle=\varepsilon_a^N\,\varphi_a^{N}({\bf r}).
\end{equation}
The diagrammatic interpretation of the various terms is straightforward as shown on fig. 1. Each contribution corresponds to the opening of the fermion lines
in the Fock energy diagram. The ordinary non local piece of the single particle hamiltonian simply comes from the opening 
of the fermion lines (fig. 1a) ignoring the dressing of the sigma propagator and the in-medium modification of the scalar coupling constant. These medium effects appear in the other diagrams and 
are depicted schematically on figure 1b,c,d,e. Opening the fermion loop renormalizing the scalar coupling constant (fig. 1b) generates 
the first rearrangement term. Opening the fermion lines associated with the dressing of the sigma propagator (fig. 1c,d,e) generates the second rearrangement term. 

The above formalism can  in principle be directly used for finite nucleus calculations. However important specific difficulties 
occur which are due in particular to the density dependence of the sigma mass and to the non trivial structure of the rearrangement terms. In practice this requires more study and we limit ourselves in the present paper to infinite nuclear matter. 

%%%%%%%%%%%%%%%%%%%%%%%%%%%%%%%%%%%%%%%%%%%%%%%%%%%%%%%%%%%%%%%%%%%%%%%%%%%%%%%%%%%%%%%%%%%%%%%%%%%%%%%%%%
\section{Infinite matter}\label{INF}
%%%%%%%%%%%%%%%%%%%%%%%%%%%%%%%%%%%%%%%%%%%%%%%%%%%%%%%%%%%%%%%%%%%%%%%%%%%%%%%%%%%%%%%%%%%%%%%%%%%%%%%%%%%ש
\label{sec:infinite matter}
In infinite nuclear matter, the single particle orbits are plane waves, labeled by momentum and spin indices $k=({\bf k}, s)$ for each isospin state  $N$~:
\begin{equation}
\label{eqn:normalisation}
\varphi^{N}_{k}({\bf r})=\frac{1}{\sqrt{V}}\,u({\bf k}, s)\,e^{i{\bf k}\cdot{\bf r}}\,\chi_N.
\end{equation}
The Fourier transform of the single particle potential can be introduced through~:
\begin{equation}
\left\langle {\bf r}\left|\gamma^0\,h^N\right|{\bf r}'\right\rangle=\int \frac{d{\bf k}}{(2\pi)^3}
\,e^{i\,{\bf k}\cdot({\bf r}\,-\,{\bf r}'})\,
\gamma^0\,h^N({\bf k})
\end{equation}
and has always the general form
\begin{equation}
\gamma^0\,h^N({\bf k})=M_N\,+\Sigma_S\left({\bf k}\right)\,+\,\vec{\gamma}\cdot{\bf k}\bigg(1+\,\Sigma_V\left({\bf k}\right)\bigg)\,+\,
\gamma^0\,\,\Sigma_0\left({\bf k}\right)
\end{equation}
where $\Sigma_S\left({\bf k}\right)$, $\Sigma_V\left({\bf k}\right)$ and $\Sigma_0\left({\bf k}\right)$ are  the various self-energies of scalar or vector nature. Notice that the above quantities are {\it a priori} different for protons and neutrons.
The Hartree-Fock equation becomes :
\begin{equation}
h^N({\bf k})u({\bf k}, s)=\varepsilon_a^Nu({\bf k}, s).
\end{equation}
This equation is formally identical to the Dirac equation for a free particle, giving for our normalisation (\ref{eqn:normalisation}):
\begin{equation}
u({\bf k}, s)=\sqrt{\frac{E^*+M^*}{2E^*}}\left(
\begin{array}{c}
                \chi \\
                \frac{\bfvec{\sigma}\cdot {\bf k}^*}{E^*+M^*}\chi
\end{array}
\right)
\end{equation}
where
\begin{eqnarray}
\label{eqn:effectivequantities}
E^* &=& \varepsilon_a^N-\Sigma_0\left({\bf k}\right) = \sqrt{M^{*2}+{\bf k}^{*2}} \nonumber \\
M^* &=& M_N\,+\,\Sigma_S\left({\bf k}\right) \\
{\bf k}^* &=& {\bf k}\,\left(1\,+\,\Sigma_V\left({\bf k}\right)\right) \nonumber
\end{eqnarray}
are the effective energy, Dirac effective mass and effective momentum. Again notice that $M^*$ and $E^*$ are different for protons and neutrons. 
In the following formula we will distinguish them with $p$ and $n$ indices. 
We also introduce the Fermi momenta, $p_F$ for the protons  and $n_F$ for the neutrons, the occupation numbers for the protons,
$N_{p{\bf k}}=\Theta(p_F\,-\,|{\bf k})|$, and for the neutrons $N_{n{\bf k}}=\Theta(n_F\,-\,|{\bf k})|$ and 
the vector and scalar density for protons and neutrons~:
\begin{eqnarray}
& &\rho_p=\int{\frac{2\,d{\bf k}}{(2\pi)^3}\,N_{p{\bf k}}},\qquad\quad \rho_n=\int{\frac{2\,d{\bf k}}{(2\pi)^3}\,N_{n{\bf k}}},
\qquad\quad\rho=\rho_p\,+\,\rho_n\nonumber\\
& &\rho_{Sp}=\int{\frac{2\,d{\bf k}}{(2\pi)^3}\,N_{p{\bf k}}\,\frac{M^*_p}{E^*_p}},\quad 
\rho_{Sn}=\int{\frac{2\,d{\bf k}}{(2\pi)^3}\,N_{n{\bf k}}\,\frac{M^*_n}{E^*_n}},\quad\rho_S=\rho_{Sp}\,+\,\rho_{Sn}.
\end{eqnarray}
The $S_p$ and $S_n$ matrices take the explicit form~:
\begin{eqnarray}
\big(S_p({\bf r}-{\bf r}')\big)_{\alpha\beta}&=& \frac{1}{V}\sum_{{\bf k}}\,e^{i{\bf k}\cdot({\bf r}-{\bf r}')}
\left(\frac{\fslash{{k}}^*+M^*}{2E^*}\right)_{\alpha\beta,p}\,N_{p{\bf k}}\nonumber\\
\big(S_n({\bf r}-{\bf r}')\big)_{\alpha\beta}&=&\frac{1}{V}\sum_{{\bf k}}\,e^{i{\bf k}\cdot({\bf r}-{\bf r}')}
\left(\frac{\fslash{{k}}^*+M^*}{2E^*}\right)_{\alpha\beta,n}\,N_{n{\bf k}};
\end{eqnarray}
The energy density can be written as~:
\begin{equation}
\epsilon=\epsilon_{kin+Hartree}\,+\,\epsilon_{Fock}
\end{equation}
Once the equations  of motion for the expectation value of the classical $\omega_0$, $\rho_0$ and $\delta$ fields have been used, the Kinetic+Hartree contribution to the energy density has the form~:
\begin{eqnarray}
\epsilon_{kin+Hartree}&=& \int \frac{2\,d{\bf k}}{(2\pi)^3}\,\sum_N\,\left({\bf k}\cdot\frac{{\bf k}^*}{E^*}\,+\,M_N(\bar s)\,\frac{M^*}{E^*}\right)_N
\,+\,V(\bar s) \nonumber \\
& & +\frac{1}{2}\left(\frac{g_\omega}{m_\omega}\right)^2\rho^2-\frac{1}{2}\left(\frac{g_\delta}{m_\delta}\right)^2(\rho_{sp}-\rho_{sn})^2+\frac{1}{2}\left(\frac{g_\rho}{m_\rho}\right)^2(\rho_p-\rho_n)^2.\label{EPSKINHA}
\end{eqnarray}
The Fock contribution from scalar exchange is given by~:
\begin{equation}
\epsilon_{Fock}^{(s)} = \frac{g_S^{*2}}{2} \int \frac{d{\bf k}}{(2\pi)^3}\frac{d{\bf k}'}{(2\pi)^3}
\frac{1}{({\bf k}-{\bf k}')^2+ m_\sigma^{*2}}\sum_{N}
\left(1+\frac{M^*}{E^*}\frac{M^{'*}}{E^{'*}}-\frac{{\bf k}^*}{E^*}\cdot\frac{{\bf k}^{'*}}{E^{'*}}\right)_N\,N_{\bf k}\,N_{{\bf k}'}
\label{FOCKENERGY}
\end{equation}
and the other Fock terms are listed in appendix B.\\

Let us now come to the various contributions to the self-energy. Ignoring for the moment the rearrangement terms, they can be obtained starting 
from eq. (\ref{ORDSE}). 
For the protons the scalar component of the self-energy is~:
\begin{eqnarray}
\Sigma_S\left({\bf k}\right) &=& M_N(\bar{s})\,-\,M_N\,-\,\frac{g^2_\delta}{m^2_\delta}\,(\rho_{Sp}\,-\,\rho_{Sn}) \nonumber \\
& & +\,\sum_{R,R';N'}\frac{g_{R R'}^{(p N')}}{2}\int\frac{d{\bf k}'}{(2\pi)^3}\frac{1}{({\bf k}-{\bf k'})^2+m_R^2}\,\frac{1}{4}Tr\bigg(\Gamma_R\,\frac{\fslash{{k}}'^*+M'^*}{2E'^*}\,\Gamma_{R'}\bigg)\,N_{{\bf k}'}\nonumber\\
& & +\,\sum_{R,R';N'}\frac{g_{R R'}^{(p N')}}{2}\int\frac{d{\bf k}'}{(2\pi)^3}\frac{1}{({\bf k}-{\bf k'})^2+m_R^2}\,\frac{1}{4}Tr\bigg(\Gamma_{R'}\,\frac{\fslash{{k}}'^*+M'^*}{2E'^*}\,\Gamma_{R}\bigg)\,N_{{\bf k}'},
\label{SELFS}
\end{eqnarray}
the time-component part of the vector self-energy is ~:
\begin{eqnarray}
\Sigma_0\left({\bf k}\right) &=& \frac{g^2_\omega}{m^2_\omega}\,(\rho_p\,+\,\rho_n)\,+\,\frac{g^2_\rho}{m^2_\rho}\,(\rho_p\,-\,\rho_n) \nonumber \\
& & +\,\sum_{R,R';N'}\frac{g_{R R'}^{(p N')}}{2}\int\frac{d{\bf k}'}{(2\pi)^3}\frac{1}{({\bf k}-{\bf k'})^2+m_R^2}\,\frac{1}{4}Tr\bigg(\gamma_0\,\Gamma_R\,\frac{\fslash{{k}}'^*+M'^*}{2E'^*}\,\Gamma_{R'}\bigg)\,N_{{\bf k}'}\nonumber\\
& & +\,\sum_{R,R';N'}\frac{g_{R R'}^{(p N')}}{2}\int\frac{d{\bf k}'}{(2\pi)^3}\frac{1}{({\bf k}-{\bf k'})^2+m_R^2}\,\frac{1}{4}Tr\bigg(\gamma_0\,\Gamma_{R'}\,\frac{\fslash{{k}}'^*+M'^*}{2E'^*}\,\Gamma_{R}\bigg)\,N_{{\bf k}'}
\label{SELFVT}
\end{eqnarray}
and the spatial part  is~:
\begin{eqnarray}
\Sigma_V\left({\bf k}\right) &=& 
\sum_{R,R';N'}\frac{g_{R R'}^{(p N')}}{2}\int\frac{d{\bf k}'}{(2\pi)^3}\frac{1}{({\bf k}-{\bf k'})^2+m_R^2}\,\frac{1}{4}
Tr\bigg(\vec\gamma\cdot\hat{\bf k} \,\Gamma_R\,\frac{\fslash{{k}}'^*+M'^*}{2E'^*}\,\Gamma_{R'}\bigg)\,N_{{\bf k}'}\nonumber\\
& +& \,\sum_{R,R';N'}\frac{g_{R R'}^{(p N')}}{2}\int\frac{d{\bf k}'}{(2\pi)^3}\frac{1}{({\bf k}-{\bf k'})^2+m_R^2}\,\frac{1}{4}Tr\bigg(\vec\gamma\cdot\hat{\bf k}\,\Gamma_{R'}\,\frac{\fslash{{k}}'^*+M'^*}{2E'^*}\,\Gamma_{R}\bigg)\,N_{{\bf k}'}.
\label{SELFVS}
\end{eqnarray}
For the neutrons, the self-energies are the same as those for the protons except that the $p$ and $n$ indices are exchanged.
Note that in  infinite nuclear matter, the expectation value $\bar s$ does not depend on the position, therefore the Fourier transform of the $s$
 propagator  is as simple as for the other mesons, simply replacing the vacuum mass by the in-medium sigma mass.
Here we have decomposed the $\rho$ term into a vector one, a tensor one and a crossed one between the vector and the tensor interactions. 

Finally, the rearrangement terms give  momentum independent contributions to the scalar self energy~:
\begin{equation}
\Sigma_{S\,(rg\,1)}=-\,2\,\frac{\tilde{\kappa}_{NS}}{m^{*2}_{\sigma}}\,\epsilon_{Fock}^{(s)}
\end{equation}
\begin{eqnarray}
\Sigma_{S\,(rg\,2)}&=&	-\frac{\partial m^{*2}_{\sigma}}{\partial \rho_S}\,\frac{g_S^{*2}}{2}\, \int \frac{d{\bf k}}
{(2\pi)^3}\frac{d{\bf k}'}{(2\pi)^3}
\left(\frac{1}{({\bf k}-{\bf k}')^2+m_\sigma^{*2}}\right)^2\nonumber\\
& &\qquad\sum_{N}\left(1+\frac{M^*}{E^*}\frac{M^{'*}}{E^{'*}}-\frac{{\bf k}^*}{E^*}\cdot\frac{{\bf k}^{'*}}{E^{'*}}\right)_N\,N_{\bf k}\,N_{{\bf k}'}
\end{eqnarray}
where the derivative of the in-medium sigma mass with respect to the scalar density is~:
\begin{eqnarray}
\frac{\partial m^{*2}_{\sigma}}{\partial\rho_S}&=&	\left(V'''(\bar s)\,+\,\frac{\partial\tilde{\kappa}_{NS}}{\partial\bar s}_,\rho_S\right)\,
\frac{\partial\bar s}{\partial\rho_S}\,+\,\tilde{\kappa}_{NS}\nonumber\\
&=&
\tilde{\kappa}_{NS}\,-\,\frac{g^{*}_{S}}{m^{*2}_{\sigma}}\left(V'''(\bar s)\,+\,\frac{\partial\tilde{\kappa}_{NS}}{\partial\bar s}
\,\rho_S\right).
\end{eqnarray}
%%%%%%%%%%%%%%%%%%%%%%%%%%%%%%%%%%%%%%%%%%%%%%%%%%%%%%%%%%%%%%%%%%%%%%%%%%%%%%%%%%%%%%%%%%%%%%%%%%%%%
\section{Results}
%%%%%%%%%%%%%%%%%%%%%%%%%%%%%%%%%%%%%%%%%%%%%%%%%%%%%%%%%%%%%%%%%%%%%%%%%%%%%%%%%%%%%%%%%%%%%%%%%%%%%
\label{RES}
In the previous sections  the results for the energy (eq. \ref{FOCKENERGY}) and the self-energy (eq. \ref{SELFS},\ref{SELFVT},\ref{SELFVS})
have been  presented in a way which  exhibits the structure of the interactions and  in particular the explicit form of the meson propagators. However in practice  the angular integrations relative to the angle between  ${\bf k}$ and ${\bf k}'$ are performed analytically and we obtain results which are similar in shape to the one used in ref. \cite{BMVM87} for the numerical calculations.

As explained before the spirit of this paper is  to study nuclear matter properties with parameters fixed as most as possible by lattice data and hadronic phenomenology. For the well established masses we take $M_N=938.9$ MeV, $m_\omega=783$ MeV, $m_\delta=984.7$ MeV,
$m_\pi=139.6$ MeV and  $m_\rho=779$ MeV. The pseudo-vector coupling constant of the pion is $g_A/(2f_\pi)$ where $g_A=1.25$ is the axial coupling constant  and $f_\pi=94$ MeV is the pion decay constant. For the rho meson coupling constant we take the VDM value $g_\rho=2.65$. The scalar coupling constant in the vacuum is fixed to the linear sigma model value $g_S=10$. From a comparison with lattice data we deduced a sigma mass 
$m_\sigma=800$ MeV. Hence the only parameters are the scalar suceptibility $K_{NS}$ (or the dimensionless parameter 
$C=(f_\pi^2/2M_N)\kappa_{NS}$)  and the omega-nucleon coupling constant, $g_\omega$. Nevertheless we do not take them as really free parameters but only allow small deviations around the lattice estimate $C\simeq 1.25$ and the VDM/quark model value $g_\omega=3\times  2.65\simeq 8$.

One uncertainty which is particularly important for the asymmetry energy, $a_S$, is the tensor coupling constant of the rho to the nucleon. We will consider different cases between the pure VDM value $\kappa_\rho=3.7$ and the value deduced from scattering data $\kappa_\rho=6.6$ \cite{HP75}. For each case the susceptibility parameter $C$ and $g_\omega$ will be fixed.

Finally we remark, as the Orsay group \cite{BMVM87,BERNARD93}, that the pion exchange and the tensor part of the rho exchange NN interaction  contain a momentum independent piece which corresponds in configuration space to a contact interaction. In realistic many body calculations 
this $\delta({\bf r})$ contribution has to be suppressed by short-range correlations. The prescription of the Orsay group, also used in \cite{GMST06,RGMT07}, was simply to remove the contact term, {\it i.e.}, making the replacement in the Fock exchange terms~:
\begin{eqnarray}
\frac{{\bf q}^2}{{\bf q}^2\,+\,m_\pi^2}&\to &\frac{{\bf q}^2}{{\bf q}^2\,+\,m_\pi^2}\,-\,1=-\,\frac{m_\pi^2}{{\bf q}^2\,+\,m_\pi^2}\nonumber\\
\frac{{\bf q}^2}{{\bf q}^2\,+\,m_\rho^2}&\to &\frac{{\bf q}^2}{{\bf q}^2\,+\,m_\rho^2}\,-\,1=-\,\frac{m_\rho^2}{{\bf q}^2\,+\,m_\rho^2}.
\end{eqnarray}
In other words, the pion and tensor rho exchange become ordinary Yukawa potentials and they give an attractive contribution to the energy per particle. In the following we will refer this prescription as ``without contact term''. We will also consider a more realistic case where the pion and rho exchanges are folded with a two body correlation function $G({\bf r})$ \cite{W77,CDE85}. As an example we take $G(r)=j_0(q_C\,r)$ ($q_C=m_\omega$). This is equivalent to make a new change~:
\begin{eqnarray}
-\,\frac{m_\pi^2}{{\bf q}^2\,+\,m_\pi^2}&\to &-\,\frac{m_\pi^2}{{\bf q}^2\,+\,m_\pi^2}+\,\frac{m_\pi^2}{{\bf q}_C^2\,+\,m_\pi^2}\nonumber\\
-\,\frac{m_\rho^2}{{\bf q}^2\,+\,m_\rho^2}&\to &-\,\frac{m_\rho^2}{{\bf q}^2\,+\,m_\rho^2}+\,\frac{m_\rho^2}{{\bf q}_C^2\,+\,m_\rho^2}.
\label{CONTACT}
\end{eqnarray}
In conventionnal many-body language this is also equivalent to include a Landau-Migdal spin-isospin central interaction characterized by a $g'$ parameter \cite{W77,CDE85}~:
\begin{equation}
	g'=\frac{1}{3}\,\frac{{\bf q}_C^2}{{\bf q}_C^2\,+\,m_\pi^2}\,+\,
\frac{2}{3}\,C_\rho\,\frac{{\bf q}_C^2}{{\bf q}_C^2\,+\,m_\rho^2}	\qquad\hbox{with}\qquad
C_\rho=\left(\frac{f_\pi\,g_\rho\,\kappa_\rho}{g_A\,M_N}\right)^2
\end{equation}
Taking the pure VDM value, $\kappa_\rho=3.7$, one obtains $g'=0.53$, whereas the strong rho case, $\kappa_\rho=6.6$, leads to 
$g'=0.95$. These values have to be compared with the  most recent analysis which give $g'\simeq 0.6$ \cite{ISW06}. Since in this case it remains a contact term we will refer it  as ``with contact term''. Notice that the case ``without contact term'' can be recovered by taking the limit $q_C\to \infty$; this corresponds to $g'=0.73$ ($\kappa_\rho=3.7$) and $g'=1.7$ ($\kappa_\rho=6.6$).

%%%%%%%%%%%%%%%%%%%%%%%%%%%%%%%%%%%%%%%%%%%%%%%%%%%%%%%
\subsection{Infinite symmetric matter}
%%%%%%%%%%%%%%%%%%%%%%%%%%%%%%%%%%%%%%%%%%%%%%%%%%%%%%%%%%%%%%%%%%%%%%%%%
We vary the $\omega NN$ coupling constant   and the value of $C$ in the (small) parameter region described above to reproduce the   saturation density and the binding energy of nuclear matter : $\rho_0=0.16$ fm$^{-3}$ and $E/A(\rho_0)=-15.96$ MeV. Another crucial parameter (although less well fixed)   is the compressibility modulus $K=9\rho^2\partial^2\epsilon/\partial\rho^2$ which should be around $250$MeV. 
As we will  see, the present approach have difficulties to obtain such a low value for  $K$ and we choose  to first reproduce the saturation density $\rho_0$ and the binding energy $E/A$. 

Our results are given in Table 1 and the corresponding saturation curves are shown on fig. 2. For this discussion we keep the rho tensor coupling to its VDM value, namely $\kappa_\rho=3.7$. We give our results with or without the dependence of $\tilde\kappa_{NS}$ on $\bar s$ and with or without the contact interaction. The first two lines  of the table correspond to values of the parameters, ($g_\omega$, $C$), giving a correct saturation point. In the last line of the table we give  the asymmetry energy parameter $a_S$, anticipating the next subsection.
\begin{table}[h]
\begin{center}
\begin{tabular}{|l|c|c|c|c|}
\hline
& \multicolumn{2}{c|}{Without contact} & \multicolumn{2}{c|}{With contact} \\
& $\kappa_{NS}$ & $\tilde\kappa_{NS}(\bar s)$ & $\kappa_{NS}$ & $\tilde\kappa_{NS}(\bar s)$ \\
\hline \hline
$g_\omega$ & 7.775 & 7.678 & 6.52 & 6.42 \\
\hline
C & 1.33 & 1.46 & 1.49 & 1.62 \\
\hline \hline
$\rho/\rho_0$ & 1.00 & 1.00 & 1.00 & 1.00 \\
\hline
E/A(MeV) & -15.96 & -15.93 & -15.97 & -15.92 \\
\hline
K(MeV) & 316 & 297 & 298 & 281 \\
\hline \hline
$a_S$(MeV) & 29.58 & 29.45 & 26.71 & 26.64 \\
\hline
\end{tabular}
\end{center}
\label{tab:infinitematter}
\caption{Values of the parameters and coordinates of the saturation point for different cases in symmetric matter. The last line is the asymmetry energy. The rho tensor coupling is $\kappa_\rho=3.7$ (VDM value). The labels ``without contact'' and ``with contact'' interactions  are explained in the text. The colums with $\tilde\kappa_{NS}(\bar s)$ correspond to the case where the scalar susceptibility is density dependent and vanishes at full restoration.}
\end{table}
\begin{figure}[h]
\begin{center}
\includegraphics[scale=0.5,angle=270]{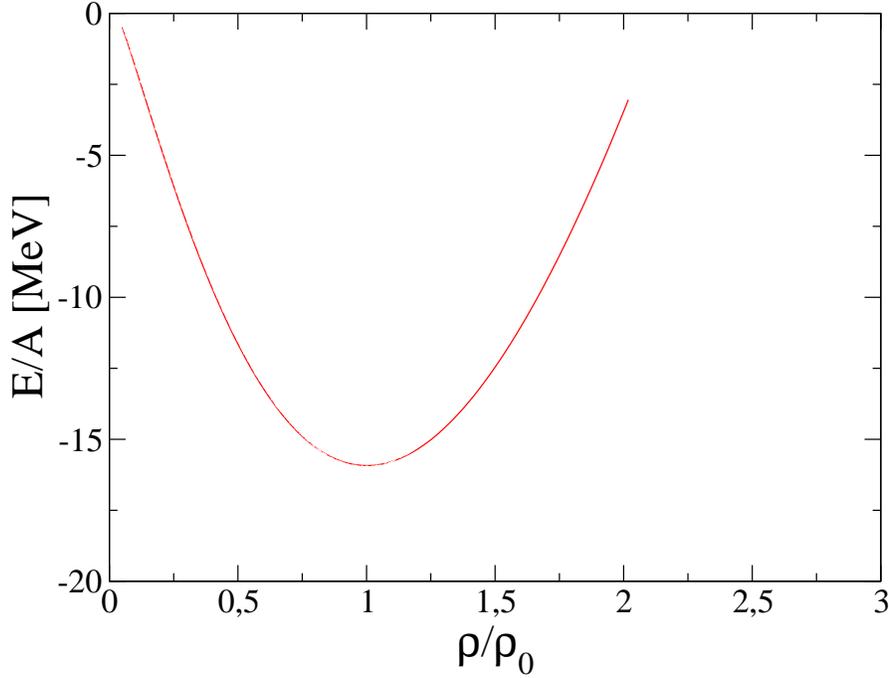}
\end{center}
\label{fig:saturation}
\caption{Saturation curves for the cases  including the contact terms with (solid) or without (dotted-dashed) the dependence of $\tilde\kappa_{NS}$ on $\bar s$ and cases without the contact terms with (dotted) or without (dashed) the dependence of $\tilde\kappa_{NS}$ on $\bar s$.}
\end{figure}
\begin{figure}[h]
\begin{center}
\includegraphics[scale=0.5,angle=0]{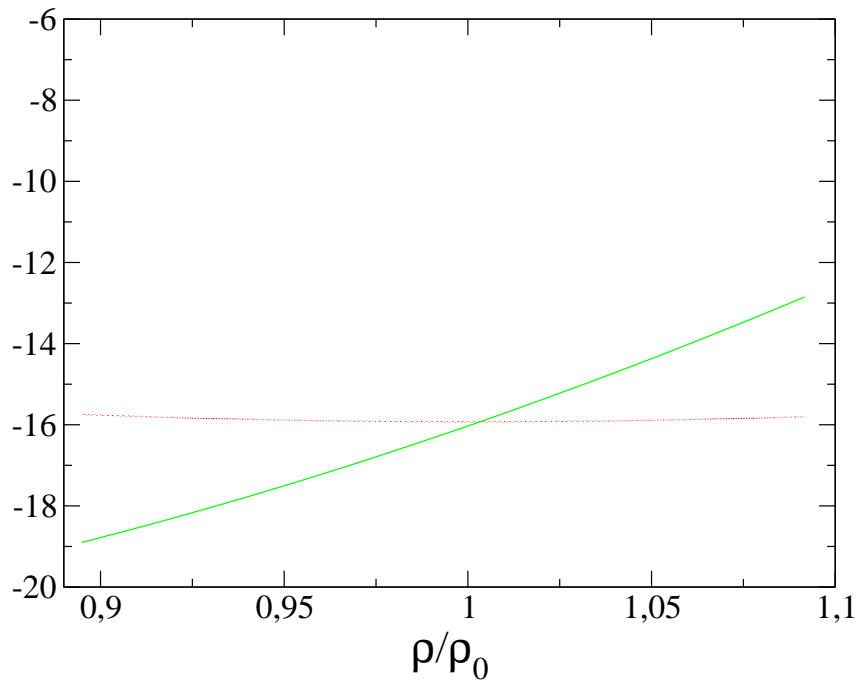}
\end{center}
\label{fig:chempot}
\caption{Chemical potential  with (solid line) or without (dashed line ) inclusion of the rearrangement terms. For comparison, the energy per particle (dotted line) is also shown.}
\end{figure}
\begin{figure}[h]
\begin{center}
\includegraphics[scale=0.5,angle=0]{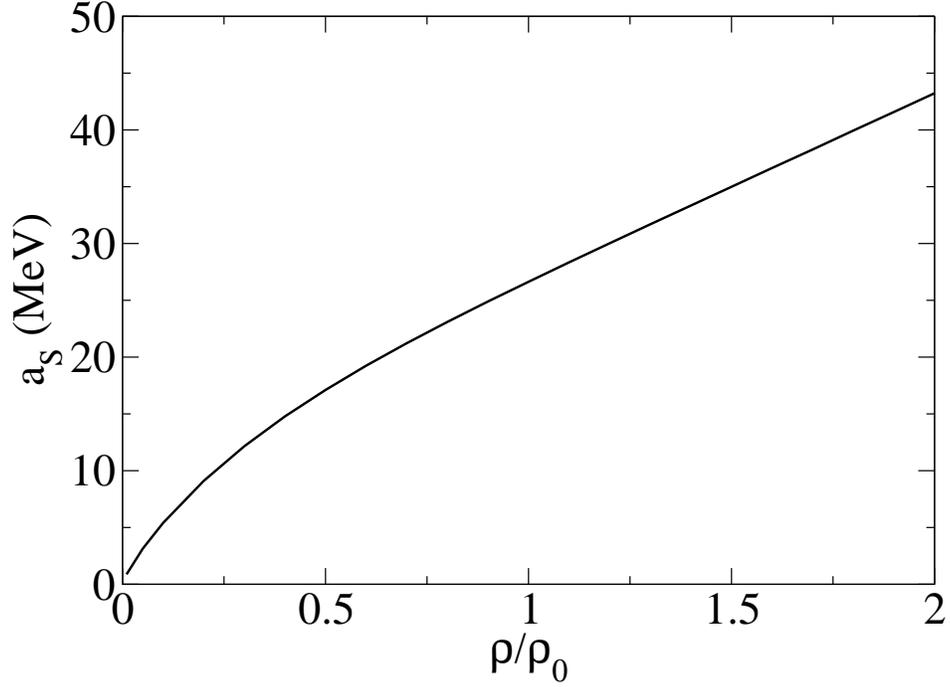}
\end{center}
\label{fig:asy-1}
\caption{Asymmetry energy versus density. The calculation is done with the inclusion of the contact terms and when  the scalar polarizability $\tilde\kappa_{NS}$ depends on $\bar s$. }
\end{figure}
What come out from these calculations 
is a globally  too large value of the compressibility modulus $K$ even there is some numerical uncertainty  in its determination and even if its extraction is not free of ambiguity in the context of relativistic theories \cite{CV04}. Removing the contact term ({\it i.e.,} taking the Orsay prescription for the treatment of the pion and tensor rho exchanges) has the advantage to reduce the $C$ parameter to a value closer to the lattice estimate,
$C\simeq 1.25$,  but the incompressibillity  $K$ still increases by about  $15-20$ MeV.  These contact terms in pion and rho exchanges (second term of the l.h.s of eq. \ref{CONTACT}) both give a repulsive contribution linear in density to the binding energy per nucleon: one finds 
$11.85$ MeV for the rho contact piece and  $0.6$ MeV for the pion contact piece and the net attractive Fock term of the rho ($-10.8$ MeV) and the pion ($-5$ MeV) are thus reduced. Since these contact terms behave like an effective omega exchange, it is natural that  the needed $\omega NN$ coupling constant is reduced  when they are incorporated as apparent on Table 1. The  numbers relative to the Fock pion and rho exchange given just above correspond to the last column of Table 1 but they are actually very similar in the other cases.    Incorporating the density dependence of the scalar susceptibility ($\tilde\kappa_{NS}(\bar s)$ columns) has approximatively  the same effect as the absence of the contact terms in the sense that one has to increase the $C$ parameter to compensate the lost repulsion.

The main effect of the rearrangement terms is  to  reduce the effective nucleon mass by a few MeV but their  influence on the saturation point is essentially negligible. This feature can be easily understood since these rearrangement terms enter the binding energy only through the dependence of the single particle wave functions on the Dirac effective  mass $M^*=M_N\,+\,\Sigma_S$ and in the non relativistic limit this dependence just disappears.  These  rearrangement terms have nevertheless an important  influence on the chemical potential, which is defined by :
\begin{equation}
\mu=\frac{d\epsilon}{d\rho}=E^*(k_F)+\Sigma^0(k_F).
\end{equation}
The Hugenholtz-Van-Hove (HVH) theorem \cite{HVH58} states that the chemical potential is equal to the binding energy per particle at the saturation point. On figure 3, we show the result of the calculation   for the chemical potential $\mu$ and for the binding energy per particle versus density, calculated again  with the contact terms and with the density dependence of $\tilde\kappa_{NS}$. We show the values of $\mu$ with or without the rearrangement terms, the former being a few MeV smaller due the their  effect  on $E_F$. We see that in the presence of rearrangement terms, the theorem is exactly  satisfied. Omitting the rearrangement terms yields to a significant violation of the HVH theorem. This failure has only a very moderate effect
for the saturation curve of  nuclear matter but would lead to major problems in finite nuclei, since
the position of the Fermi energy would be displaced by about $5$ MeV.

As already mentionned, in the non relativistic limit the orbital wave functions are imposed: these are simply non relativistic plane waves. Consequently if the system is approximatively non relativistic, the results of the calculations should not depend very much on the choice of the wave functions. Said differently, the results obtained with  another basis than the fully self-consistent Hartree-Fock (HF) basis  would deviate from the HF results only by relativistic effects. For this reason we introduce the Hartree basis for which the Fock terms appearing in the self-energies (eq. \ref{SELFS},\ref{SELFVT},\ref{SELFVS}) are removed. One advantage is the strong simplification of the calculations.
The results for the saturation parameters are displayed on table 2. As expected, the results are very similar but 
the results for  $C$ and for the asymmetry energy are slightly better in the Hartree-Fock basis. However the compressibility modulus 
decreases when  the Hartree basis is used.
\begin{table}[h]
\begin{center}
\begin{tabular}{|l|c|c|c|c|}
\hline
& \multicolumn{2}{c|}{Without contact} & \multicolumn{2}{c|}{With contact} \\
& $\kappa_{NS}$ & $\tilde\kappa_{NS}(\bar s)$ & $\kappa_{NS}$ & $\tilde\kappa_{NS}(\bar s)$ \\
\hline \hline
$g_\omega$ & 7.728 & 7.615 & 6.45 & 6.36 \\
\hline
C & 1.47 & 1.62 & 1.61 & 1.735 \\
\hline \hline
$\rho/\rho_0$ & 1.00 & 1.00 & 0.99 & 1.00 \\
\hline
E/A(MeV) & -15.96 & -15.96 & -16.01 & -15.98 \\
\hline
K(MeV) & 301 & 284 & 286 & 274 \\
\hline \hline
$a_{sym}$(MeV) & 26.15 & 26.13 & 24.10 & 24.11 \\
\hline
\end{tabular}
\end{center}
\label{tab:hartree}
\caption{The same as Table 1 but  in the Hartree basis.}
\end{table}
%%%%%%%%%%%%%%%%%%%%%%%%%%%%%%%%%%%%%%%%%%%%%%%%%%%%%%%%%%%%%%%%%%%%%%%%
\subsection{Asymmetric matter}
%%%%%%%%%%%%%%%%%%%%%%%%%%%%%%%%%%%%%%%%%%%%%%%%%%%%%%%%%%%%%%%%%%%%%%%%

The properties of asymmetric nuclear matter also constitute a crucial test for any effective theory. This is particularly true for the study of nuclei far from stability and  the equation of state of asymmetric matter is an important input in astrophysical studies and in particular for neutron stars properties. For that purpose let us  introduce the asymmetry (isospin)  parameter:
$\beta=(\rho_n-\rho_p)/\rho$.

\begin{table}[h]
\begin{center}
\begin{tabular}{|l|c|c|c|}
\hline
%& \muticolumn{2}{c|}{Without contact} & \muticolumn{2}{c|}{With contact} \\
$\kappa_\rho$ & 3.7 & 5 & 6.6 \\
\hline \hline
$g_\omega$ & 6.42 & 7.348 & 8.62 \\
\hline
C & 1.62 & 1.32 & 0.93 \\
\hline \hline
$\rho/\rho_0$ & 1.00 & 1.00 & 1.00 \\
\hline
E/A(MeV) & -15.92 & -16.05 & -15.94 \\
\hline
K(MeV) & 281 & 313 & 308 \\
\hline \hline
$a_{sym}$(MeV) & 26.64 & 29.77 & 35.87 \\
\hline
\end{tabular}
\end{center}
\label{tab:asymmetry}
\caption{Evolution of the asymmetry energy with the value of the $\rho$-tensor coupling to the nucleon. The calculation is done with the inclusion of the contact terms and when  the scalar polarizability $\tilde\kappa_{NS}$ depends on $\bar s$. }
\end{table}
\bigskip\noindent
%%%%%%%%%%%%%%%%%%%%%%%%%%%%%%%%%%%%%%%%%%%%%%%%%%%%%%%%%%%%%%%%%%%%%%%%%%%%%%%%%%%%%%%%%%%%%%%%%%%%%%%%%%%%%%%%%%%%%%%%%%%%%%%%%%%
\paragraph{Asymmetry energy.} The first quantity to be considered is the asymmetry energy which can be defined according to~:
\begin{equation}
	a_S(\rho)=\left[\frac{\partial (E/A)}{\partial\beta}\right]_{\beta=0}=\left[\frac{\partial (\varepsilon/\rho)}{\partial\beta}\right]_{\beta=0}.
\end{equation}
The contribution of the kinetic+Hartree term (eq. \ref{EPSKINHA}) can be  calculated analytically. In a pure Hartree approximation, omitting the extremely small contribution from the $\delta$ meson 
exchange and the very tiny dependence of the scalar density $\rho_S$ on $\beta$, one  obtains the familiar result~:
$$ (a_S)_{kin+Hartree}=\frac{k^2_F}{6\,\sqrt{k^2_F\,+\,M^{*2}}}\,+\,\frac{g^2_\rho}{2\,m^2_\rho}\,\rho.
$$
In the actual calculation (last column of Table 1) the kinetic energy term gives $(a_S)_{kin}=13.9$ MeV, and the Hartree rho exchange term gives  $(a_S)_{rho-Hartree}$=7.07 MeV. Hence The Hartree piece, $(a_S)_{Hartree}=21$ MeV, is not sufficient to reproduce the asymmetry energy. 
It follows that  the Fock terms are absolutely necessary to get an higher value closer to the accepted values centered around $30$ MeV. As for the binding energy there is a strong compensation 
between the scalar and omega exchanges: $(a_S)_{s+\omega}=-0.75$ MeV and the needed repulsion  comes from the pion exchange
$(a_S)_{pion}=2.75$ MeV and mainly from the rho exchange $(a_S)_{rho-Fock}=5.33$ MeV. The important point is the specific contribution of the tensor piece of the rho exchange $(a_S)_{rho-tensor}=8.03-4.12\simeq 4$ MeV, the first number corresponding to the calculation 
within the Orsay prescription and the second ($-4.12$) coming from the contact term. The full result, $a_S=26.64$ MeV, is nevertheless  a little too low. However one has to keep in mind  that the  $\rho$-tensor piece goes like $\kappa_\rho^2$ and that 
the $\rho$-tensor tensor coupling is not very well known phenomenologically. In the above discussion we took the pure VDM value, 
$\kappa_\rho=3.7$, but the understanding of $\pi-N$ scattering data requires $\kappa_\rho=6.6$ \cite{HP75}. Given this uncertainty we decided to vary this coupling between these two limiting values (See Table 3). For $\kappa_\rho=5$ we find  $a_S\simeq 30$ MeV. The new fit of the saturation point gives a value of the nucleon scalar response parameter, $C=1.32$,  very close to the lattice estimate but the weak point is that the compressibilty modulus increases beyond $300 MeV$. Independently of the details of the model calculation the above discussion strongly suggests that  this $\rho$-tensor Fock term, which is absent in RMF theory, is a crucial ingredient to get a consistent description of bulk properties of symmetric and asymmetric nuclear matter. Its importance has also been stressed in the framework of density-dependent relativistic Hartree-Fock theory (DDRHF) where it is shown that it significantly improves the results for the single particle spectra in finite nuclei \cite{LSVJ07}. For completeness we also plot on fig. 4 the asymmetry energy against 
the density. We see a monotonic increase of $a_S$ but with a slope decreasing at high density, which is very similar in shape to the result of a BHF calculation \cite{ZBL99}. In any case this curve never exhibits the  maximum found in the first version of in-medium chiral perturbation calculation \cite{KFW02}.
\begin{figure}[h]
\begin{center}
\includegraphics[scale=0.5,angle=270]{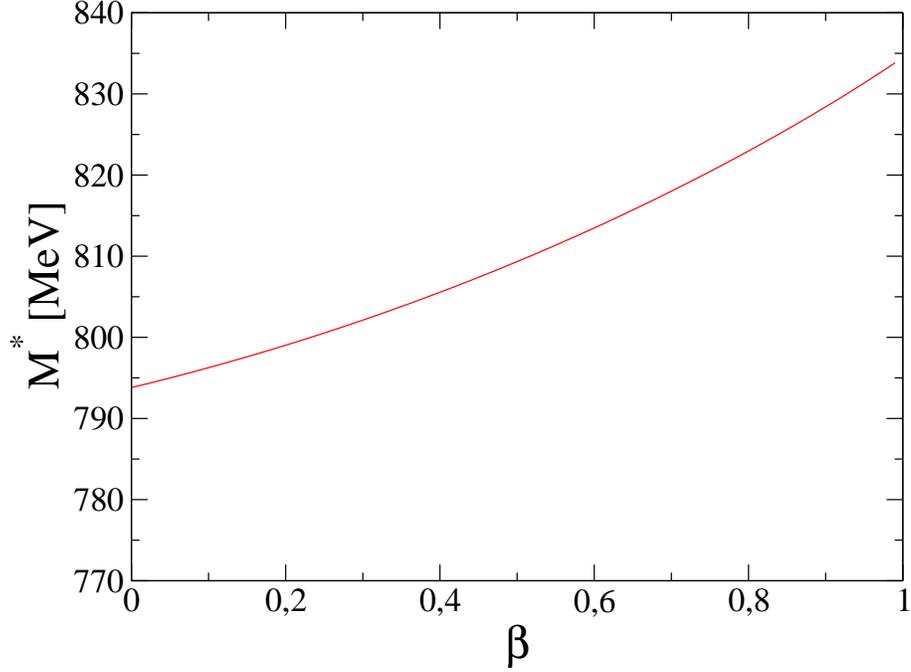}
\end{center}
\label{fig:landaumass}
\caption{Landau masses for neutrons (solid line ) and protons (dashed line ) in neutron-rich matter at baryonic density $\rho_0$. The calculation is done with the inclusion of the contact terms and when  the scalar polarizability $\tilde\kappa_{NS}$ depends on $\bar s$. }
\end{figure} 
\bigskip\noindent
%%%%%%%%%%%%%%%%%%%%%%%%%%%%%%%%%%%%%%%%%%%%%%%%%%%%%%%%%%%%%%%%%%%%%%%%%%%%%%%%%%%%%%%%%%%%%%%%%%%%%%%%%%%%%%%%%%%%%%%%%%%%%%%%%
\paragraph{Proton and neutron effective masses.} 

The nucleon effective mass $m^*$ is a key property characterizing the propagation of the nucleons through the nuclear medium. In very exotic systems, the isosvector dependence of this effective mass, {\it i.e.}, the mass splitting between the proton effective mass, $m^*_p$, and the neutron effective mass, $m^*_n$ with increasing asymmetry $\beta$ should play an important role. However, so far, no experimental data from finite nuclei has allowed 
a determination of the effective mass splitting but  ab-initio Brueckner-Hartree-Fock calculation \cite{BL91,ZBL99} have predicted 
$m^*_n>m^*_p$ in neutron-rich matter. This result has been confirmed by relativistic calculations within the Dirac-BHF scheme \cite{SBK05} or within the DDRHF scheme \cite{LVJ06}. Thus the sign of the splitting is rather solidly predicted although its amplitude is subject to a much greater uncertainty. This effective mass, which is a momentum dependent quantity,   is defined from the density of states according to~:
\begin{equation}
\frac{1}{m^*}=\frac{1}{k}\,\frac{de}{dk}
\end{equation}
where $e$ is the single particle energy with the bare nucleon mass subtracted. In our approach it corresponds to~: 
$$e=\varepsilon_{\bf k}^N\,-\,M_N\equiv \sqrt{(M_N\,+\,\Sigma_S\left({\bf k}\right))^2\,+\,
{\bf k}^2\,\left(1\,+\,\Sigma_V\left({\bf k}\right)\right)^2}\,+\,\Sigma_0\left({\bf k}\right)-\,M_N$$
With this definition, one finds :
\begin{equation}
m^*=\frac{E^*}{(1+\Sigma_V)^2}\equiv\frac{\sqrt{(M_N\,+\,\Sigma_S\left({\bf k}\right))^2\,+\,
{\bf k}^2\,\left(1\,+\,\Sigma_V\left({\bf k}\right)\right)^2}}{(1+\Sigma_V)^2}.\label{eqn:landaumass2}
\end{equation}
Since there are many effective masses introduced in the literature a few remarks are necessary.
\begin{enumerate}
\item In the following we will discuss this effective mass taken at the Fermi momentum. This effective mass is called the Landau mass.
\item This Landau effective mass should not be confused with the Dirac mass $M^*= M_N\,+\,\Sigma_S\left({\bf k}\right))$ introduced above. The Dirac mass usually has an opposite behaviour to the Landau mass : in  neutron-rich matter, $M^*_{Dirac(n)}<M^*_{Dirac(p)}$.
\item It corresponds to the group mass in the terminology of ref. \cite{JM89} and to the relativistic mass $M^*_R$ in the terminology of ref. 
\cite{LVJ06}. Notice that the non relativistic mass $M^*_{NR}$ introduced by these authors is not directly relevant for our discussion. 
\end{enumerate}
Again we consider the case corresponding to the last column of Table 1 but the conclusions would be very similar in the other three cases. 
These effective masses are plotted on fig. 4 which demonstrates that the behaviour characterized by
 $m^*_n > m^*_p$ also occurs in our approach. It is also important to mention that the effect of the rearrangement terms is to decrease  these
  two masses by about $7$ MeV independently of the value of the neutron-richness parameter $\beta $.
\bigskip\noindent 
%%%%%%%%%%%%%%%%%%%%%%%%%%%%%%%%%%%%%%%%%%%%%%%%%%%%%%%%%%%%%%%%%%%%%%%%%%%%%%%%%%%%%%%%%%%%%%%%%%%%%%%%%%%%%%%%%%%%%%%%%%%%%%שששששששששששששש
\paragraph{ Neutron matter and cold uniform matter in beta equilibrium.} In this paragraph we determine the equation of state of cold uniform matter  which might be relevant for studies of the interior of neutron stars. In this quite preliminary work we take a simplified point of view where  this matter is mainly made of neutrons   in beta equilibrium with protons and a gas of free electrons. Hence what is presented below should be considered only as indicative results in view of future more detailed work. The energy density of such a  matter  is therefore~:
 \begin{equation}
\epsilon=\epsilon_{nuclear\, matter}\,+\,\epsilon_{e^-}
\end{equation}
 where $\epsilon_{nuclear\, matter}$ has been calculated before and $\epsilon_{e^-}$ is
\begin{equation}
\epsilon_{e^-}=\frac{2}{V}\sum_{k_e<k_{Fe}}\sqrt{k_e^2+m_e^2}\simeq\frac{k_{Fe}^4}{4\pi^2}
\end{equation}
if we neglect the electron mass. To ensure the charge neutrality of the matter, we have to impose the equality  between the densities of protons and electrons~: $\rho_p=\rho_e$ which implies $p_F=k_{Fe}$. The grand potential and its density are given by~~:
\begin{eqnarray}
\Omega &=& E\,-\,\sum_i\mu_iN_i \nonumber\\
\omega &=& \epsilon\,-\,\mu_B\,(\rho_p\,+\,\rho_n)\,-\,\mu_e\,(\rho_e\,-\,\rho_p)\nonumber \\
       &=& \epsilon\,-\,\mu_p\,\rho_p\,-\,\mu_n\,\rho_n\,-\,\mu_e\,\rho_e 
\end{eqnarray}
wich implies $\mu_B=\mu_n=\mu_p\,+\,\mu_e$. For a given total baryonic density, the proton and neutron densities and Fermi momenta can be obtained by solving  the system of equations~:
\begin{eqnarray}
k_{Fe} &=& p_F=\varepsilon^n(n_F)-\varepsilon^p(p_F) \nonumber\\
\rho &=& \rho_p\,+\,\rho_n.
\end{eqnarray}
For different values of $\rho$, we search a solution of this system around $\rho_p\simeq 0.1\,\rho$ and $\rho_n\simeq 0.9\,\rho$. With those values of $\rho_p$ and $\rho_n$, we can calculate the energy density versus baryonic density.  We plot the resulting curve  on  fig. 6 for the case corresponding again to the last column of Table 1.
We also determine  the equation of state, namely  pressure versus  energy density. Two definitions of $P$ can be used. The thermodynamical  one is:
\begin{equation}
P = -\omega=-\epsilon\,+\,\sum_i\mu_i\rho_i=-\epsilon\,+\,\mu_n\rho
\end{equation}
where the charge neutrality condition   $\rho_p=\rho_e$ has been used. The other definition of the baryonic pressure is~:
\begin{equation}
P=\rho^2\,\frac{\partial(\epsilon/\rho)}{\partial\rho}=\rho\,\frac{\partial\epsilon}{\partial\rho}\,-\,\epsilon.
\end{equation}
Since  $\mu$ is defined by $\mu_B\equiv\partial\epsilon/\partial\rho$, the two definitions of $P$ are thus equivalent.
In the same conditions as above, we plot the pressure  versus the energy on fig. 7. As all realistic calculations we find that the pure neutron matter is unbound and the energy per particle monotonically rises with the density. The equation of state (fig. 7) is rather similar to the one obtained with the QMC model \cite{RGMT07} in the density range considered ($\rho<3\,\rho_0$). According to this latter work it is clear that beyond this density the contribution of the hyperons has to be included to make a realistic study of neutron star properties. 
\vskip 2 true cm
\begin{figure}[h]
\label{eqn:landaumass}
\begin{center}
\includegraphics[scale=0.5,angle=0]{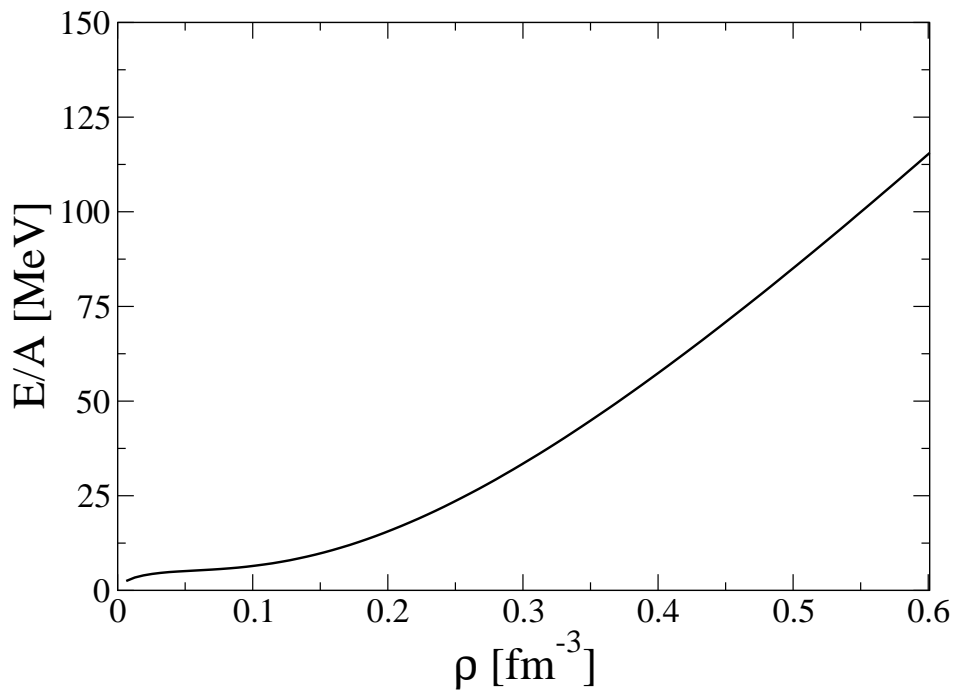}
\end{center}
\label{fig:neutronstar}
\caption{Energy per particle  versus baryonic density for cold uniform matter in beta equilibrium. The calculation is done in the scheme corresponding to the last column of Table 1.}
\end{figure}
\begin{figure}[h]
\begin{center}
\includegraphics[scale=0.5,angle=0]{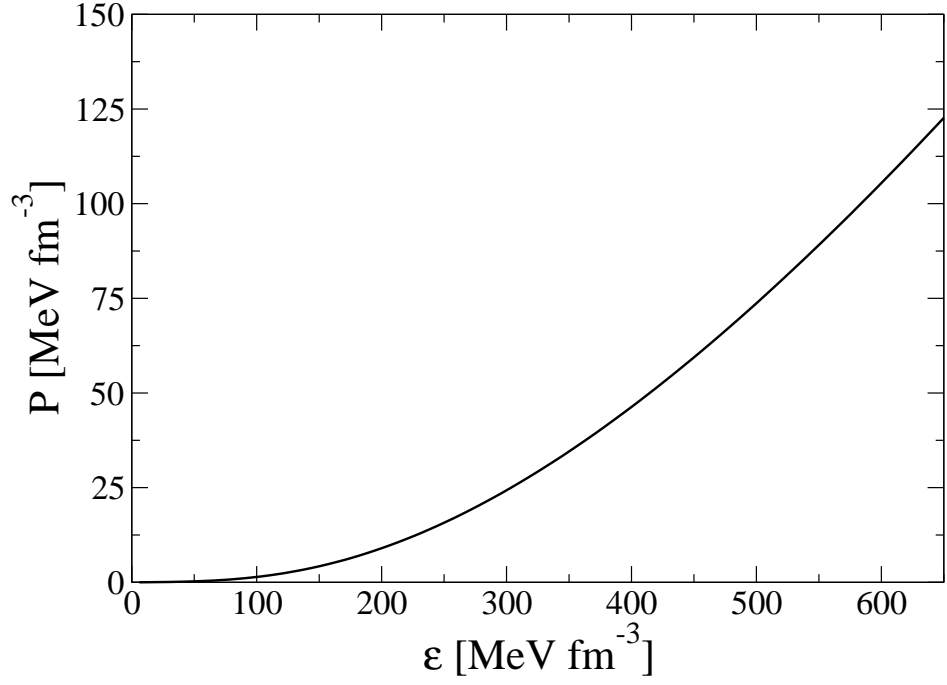}
\end{center}
\label{fig:pression}
\caption{Equation of state of cold uniform matter in beta equilibrium~: pression versus  energy density. The calculation is done in the scheme corresponding to the last column of Table 1.} 
\end{figure}

%%%%%%%%%%%%%%%%%%%%%%%%%%%%%%%%%%%%%%%%%%%%%%%%%%%%%%%%%%%%%%%%%%%%%%%%%%%%%%%%%%%%%%%%%%%%%
\section{Conclusion}
%%%%%%%%%%%%%%%%%%%%%%%%%%%%%%%%%%%%%%%%%%%%%%%%%%%%%%%%%%%%%%%%%%%%%%%%%%%%%%%%%%%%%%%%%%%%%%
In this work we have investigated the implications for the nuclear matter problem of an effective chiral relativistic theory 
where the nuclear binding is ensured by  a chiral invariant scalar field associated with the radial fluctuations of the chiral quark condensate.
Nuclear matter stability can be obtained only if the response of the nucleon to the scalar field is properly incorporated, which  phenomenologically produces   three-body repulsive forces. The saturation mechanism depends on a  balance between these repulsive forces and attractive three-body forces originating from chiral tadpole diagrams generated by the mexican hat effective potential. Once the various inputs are constrained by lattice QCD data and hadron phenomenology we obtain  encouraging results with  very minimal assumptions. One very positive point concerns the properties of asymmetric nuclear matter. We obtain the trend for the neutron-proton  Landau effective mass difference to increase with increasing $N-Z$. Similarly the asymmetry energy, $a_S$,  can be reproduced but a very important contribution comes from the tensor piece of the Fock rho exchange term. It is important to notice that the rho Hartree term is by far not sufficient if we keep the VDM value for the $\rho NN$ coupling constant. One difficulty of the approach is however a too large compressibility modulus. This is one reason among other to motivate future   works. 
The first line of investigation  is related to the many-body treatment. Here we limited ourselves to the Hartre-Fock approach in the static approximation but higher order many-body effects, such as the inclusion of the correlation  energy, have to be studied. For what concerns the pion loop correction this certainly necessitates to relax the static approximation. This inclusion of pion loops on top of the mean-field HF approach can be performed using in-medium chiral perturbation or many body approaches (RPA) in the line of our work of ref. \cite{CE07}
but extended to the relativistic case. According to our non relativistic results \cite{CE07} we  expect that this correlation energy may decrease the compressibility modulus.
Another line of research is  related to the very crucial problem of the nucleon substructure. Here we have implicitly assumed that the nucleon is made of  three constituent quarks (or possibly a constituent quark and diquark) of the NJL type linked by strings. Although very attractive this picture has some problems. For instance it seems difficult to reproduce the value of the parameter $\kappa_{NS}$ characterizing the scalar response of the nucleon \cite{EC07}. In such a picture the constituent quarks (or diquark) essentially move  in the non perturbative vaccuum. However it may very well happen that the confinement mechanism generates  a small region where the quarks are present with the highest probability. In such the region the chiral quark condensate (and consequently the quark mass) will drop  due to the presence of the valence quark scalar density. In other words the confinement mechanism may generate a "`bag"' in which chiral symmetry is restored. We are thus back to the old dilemmna of the nucleon structure. Does the nucleon look like a chirally restored bag or does it look like a Y shaped string with constituent quarks at the ends?
We believe that the scalar response of the nucleon or the chiral expansion parameter $a_4$ which are both relative to second derivatives of the nucleon mass are quantities which are very sensitive to the nucleon substructure. Certainly important efforts have to be done to obtain an estimate of these quantities for various modellings of the nucleon.
\vskip 1 true cm
{\bf Acknowledgments.} We would like to thank M. Ericson for constant interest in this work and critical reading of the manuscript. We have also benefited from discussions with the members of the nuclear structure theory group of IPNL, K. Bennaceur, T. Lesinski and J. Meyer for the questions of the nucleon effective mass and asymmetry energy. We also thank H. Hansen and M. Martini for many useful discussions.
\vfill\eject
{\bf \Large{APPENDIX}}
\appendix 

\section{Explicit form of the Fock term energy}
The various contributions to the Fock term energy, completing eq. \ref{FOCKS},   are listed below:
\begin{eqnarray}
E_{Fock}^{(s)}&=&\frac{1}{2\,}\int\,d{\bf r}\,d{\bf r}'\,Tr\big(S_p({\bf r}'-{\bf r})\,  S_p({\bf r}-{\bf r}')\,+\,
S_n({\bf r}'-{\bf r})\,  S_n({\bf r}-{\bf r}')\big)\nonumber\\
& &\qquad\qquad\qquad g^*_S({\bf r})\,g^*_S({\bf r}')\,D_\sigma({\bf r}-{\bf r}')
\end{eqnarray}
\begin{eqnarray}
E_{Fock}^{(\omega)}&=&-\frac{g^{2}_\omega}{2\,}\int\,d{\bf r}\,d{\bf r}'\,Tr\big(S_p({\bf r}'-{\bf r})\gamma^\mu  S_p({\bf r}-{\bf r}')\gamma^\nu\,+\,
S_n({\bf r}'-{\bf r})\, \gamma^\mu S_n({\bf r}-{\bf r}'\gamma^\nu)\big)\nonumber\\
& &\qquad\qquad\qquad D_{\omega\mu\nu}({\bf r}-{\bf r}')
\end{eqnarray}
\begin{eqnarray}
E_{Fock}^{(\rho)}&=&-\,\frac{g^{2}_\rho}{2\,}\int\,d{\bf r}\,d{\bf r}'\,Tr\big(S_p({\bf r}'-{\bf r})\gamma^{\mu}  S_p({\bf r}-{\bf r}')\gamma^{\nu}\,+\,
S_n({\bf r}'-{\bf r})\gamma^{\mu}  S_n({\bf r}-{\bf r}')\gamma^{\nu}\nonumber\\
& &\qquad\qquad\qquad +\, 2 \,S_p({\bf r}'-{\bf r})\gamma^{\mu}  S_n({\bf r}-{\bf r}')\gamma^{\nu}\,+\,
2\,S_n({\bf r}'-{\bf r})\gamma^{\mu}  S_p({\bf r}-{\bf r}')\gamma^{\nu})\big)\nonumber\\
& &\qquad\qquad\qquad\qquad D_{\rho\mu\nu}({\bf r}-{\bf r}')\nonumber\\
& &-2\,\,\frac{g^{2}_\rho}{2\,}\,\left(\frac{\kappa_\rho}{2\,M_N}\right)
\int\,d{\bf r}\,d{\bf r}'\,Tr\big(S_p({\bf r}'-{\bf r})\sigma^{\mu j}  S_p({\bf r}-{\bf r}')\gamma^{\nu}\nonumber\\
& &\qquad\qquad\qquad\qquad +\,
S_n({\bf r}'-{\bf r})\sigma^{\mu j}  S_n({\bf r}-{\bf r}')\gamma^{\nu}\nonumber\\
& &\qquad\qquad\qquad\qquad +\, 2 \,S_p({\bf r}'-{\bf r})\sigma^{\mu j}  S_n({\bf r}-{\bf r}')\gamma^{\nu}\,+\,
2\,S_n({\bf r}'-{\bf r})\sigma^{\mu j}  S_p({\bf r}-{\bf r}')\gamma^{\nu})\big)\nonumber\\
& &\qquad\qquad\qquad\qquad
\,\partial_j D_{\rho\mu\nu}({\bf r}-{\bf r}')\nonumber\\
& &-\,\frac{g^{2}_\rho}{2\,}\,\left(\frac{\kappa_\rho}{2\,M_N}\right)^2
\int\,d{\bf r}\,d{\bf r}'\,Tr\big(S_p({\bf r}'-{\bf r})\sigma^{\mu i}  S_p({\bf r}-{\bf r}')\gamma^{\nu j}\nonumber\\
& &\qquad\qquad\qquad\qquad +\,
S_n({\bf r}'-{\bf r})\sigma^{\mu i}  S_n({\bf r}-{\bf r}')\sigma^{\nu j}\nonumber\\
& &\qquad\qquad\qquad\qquad +\,2 \,S_p({\bf r}'-{\bf r})\sigma^{\mu i}  S_n({\bf r}-{\bf r}')\sigma^{\nu j}\,+\,
2\,S_n({\bf r}'-{\bf r})\sigma^{\mu i}  S_p({\bf r}-{\bf r}')\sigma^{\nu j})\bigg)\nonumber\\
& &\qquad\qquad\qquad\qquad \partial_i\partial'_j D_{\rho\mu\nu}({\bf r}-{\bf r}')
\end{eqnarray}
\begin{eqnarray}
E_{Fock}^{(\delta)}&=&\,\frac{g^{2}_\delta}{2\,}\int\,d{\bf r}\,d{\bf r}'\,Tr\bigg(S_p({\bf r}'-{\bf r})\,  S_p({\bf r}-{\bf r}')\,+\,
S_n({\bf r}'-{\bf r})\,  S_n({\bf r}-{\bf r}')\nonumber\\
& &\qquad+\, 2 \,S_p({\bf r}'-{\bf r})\,  S_n({\bf r}-{\bf r}')\,+\,
2\,S_n({\bf r}'-{\bf r})\,  S_p({\bf r}-{\bf r}'))\bigg)\,
D_\delta({\bf r}-{\bf r}')
\end{eqnarray}
\begin{eqnarray}
E_{Fock}^{(\pi)}&=&\,\frac{1}{2\,}\left(\frac{g_A}{2 f_\pi}\right)^2\int\,d{\bf r}\,d{\bf r}'\,Tr\bigg(S_p({\bf r}'-{\bf r})  
\,\gamma^{5}\gamma^{i}\,S_p({\bf r}-{\bf r}')\,\gamma^{5}\gamma^{j}\nonumber\\
& &\qquad\qquad +\,
S_n({\bf r}'-{\bf r})\,\gamma^{5}\gamma^{i}\,  S_n({\bf r}-{\bf r}')\,\gamma^{5}\gamma^{j}\,\nonumber\\
& &\qquad\qquad +\,2 \,S_p({\bf r}'-{\bf r})\,\gamma^{5}\gamma^{i}\,  S_n({\bf r}-{\bf r}')\,\gamma^{5}\gamma^{j}\nonumber\\
& &\qquad\qquad +_,2\,S_n({\bf r}'-{\bf r})\,\gamma^{5}\gamma^{i}\,  S_p({\bf r}-{\bf r}')\,\gamma^{5}\gamma^{j}\,)\bigg)
\,\partial_i\partial'_j D_\pi({\bf r}-{\bf r}')
\end{eqnarray}
\section{Explicit form of the energy density in infinite nuclear matter}
The various pieces  of the energy density, completing eq. (\ref{FOCKENERGY}), are listed below. For simplicity the occupation numbers,
 $N_{\bf k} N_{{\bf k}'}$,  are omitted.
\begin{eqnarray}
\epsilon_{Fock}^{(s)} &=& \frac{g_s^{*2}}{2} \int \frac{d^3{\bf k}}{(2\pi)^3}\frac{d^3{\bf k}'}{(2\pi)^3}
\frac{1}{({\bf k}-{\bf k}')^2+m_\sigma^{*2}}\sum_{N}
\left(1+\frac{M^*}{E^*}\frac{M^{'*}}{E^{'*}}-\frac{{\bf k}^*}{E^*}\cdot\frac{{\bf k}^{'*}}{E^{'*}}\right)_N \\
\epsilon_{Fock}^{(\omega)} &=& \frac{g_\omega^2}{2} \int \frac{d^3{\bf k}}{(2\pi)^3}\frac{d^3{\bf k}'}{(2\pi)^3}
\frac{1}{({\bf k}-{\bf k}')^2+m_\omega^2}\sum_{N,N'}
\left(2-4\frac{M^*}{E^*}\frac{M^{'*}}{E^{'*}}-2\frac{{\bf k}^*}{E^*}\cdot\frac{{\bf k}^{'*}}{E^{'*}}\right)_N \\
\epsilon_{Fock}^{(\delta)} &=& \frac{g_\delta^2}{2} \int \frac{d^3{\bf k}}{(2\pi)^3}\frac{d^3{\bf k}'}{(2\pi)^3}
\frac{1}{({\bf k}-{\bf k}')^2+m_\delta^2}\sum_N
\left(1+\frac{M^*}{E^*}\frac{M^{'*}}{E^{'*}}-\frac{{\bf k}^*}{E^*}\cdot\frac{{\bf k}^{'*}}{E^{'*}}\right)_N
\nonumber \\
& & +2g_\delta^2\int\frac{d^3{\bf k}}{(2\pi)^3}\frac{d^3{\bf k}'}{(2\pi)^3} \frac{1}{({\bf k}-{\bf k}')^2+m_\delta^2}
\left(1+\frac{M_p^*}{E_p^*}\frac{M_n^{'*}}{E_n^{'*}}-\frac{{\bf k}_p^*}{E_p^*}\cdot\frac{{\bf k}_n^{'*}}{E_n^{'*}}\right) \\
\epsilon_{Fock}^{(\rho^V)} &=& \frac{g_\rho^2}{2} \int \frac{d^3{\bf k}}{(2\pi)^3}\frac{d^3{\bf k}'}{(2\pi)^3}
\frac{1}{({\bf k}-{\bf k}')^2+m_\rho^2}\sum_N
\left(2-4\frac{M^*}{E^*}\frac{M^{'*}}{E^{'*}}-2\frac{{\bf k}^*}{E^*}\cdot\frac{{\bf k}^{'*}}{E^{'*}}\right)_N
\nonumber \\
& & -2g_\rho^2\int\frac{d^3{\bf k}}{(2\pi)^3}\frac{d^3{\bf k}'}{(2\pi)^3} \frac{1}{({\bf k}-{\bf k}')^2+m_\rho^2}
\left(2-4\frac{M_p^*}{E_p^*}\frac{M_n^{'*}}{E_n^{'*}}-2\frac{{\bf k}_p^*}{E_p^*}\cdot\frac{{\bf k}_n^{'*}}{E_n^{'*}}\right) \\
\epsilon_{Fock}^{(\rho^{VT})} &=& -6\frac{1}{2}g_\rho\frac{f_\rho}{2M_N} \int \frac{d^3{\bf k}}{(2\pi)^3}\frac{d^3{\bf k}'}{(2\pi)^3}
\frac{1}{({\bf k}-{\bf k}')^2+m_\rho^2}\sum_N
\left(({\bf k}-{\bf k}')\cdot\frac{{\bf k}^*}{E^*}\frac{M^{'*}}{E^{'*}}\right)_N
\nonumber \\
& & -12\frac{1}{2}g_\rho\frac{f_\rho}{2M_N} \int \frac{d^3{\bf k}}{(2\pi)^3}\frac{d^3{\bf k}'}{(2\pi)^3}
\frac{1}{({\bf k}-{\bf k}')^2+m_\rho^2}\nonumber\\
& &\qquad\times
\left(({\bf k}-{\bf k}')\cdot\frac{{\bf k}_p^*}{E_p^*}\frac{M_n^{'*}}{E_n^{'*}} 
+({\bf k}-{\bf k}')\cdot\frac{{\bf k}_n^*}{E_n^*}\frac{M_p^{'*}}{E_p^{'*}}\right) \\
\epsilon_{Fock}^{(\rho^T)} &=& \frac{1}{2}\left(\frac{f_\rho}{2M_N}\right)^2 \int 
\frac{d^3{\bf k}}{(2\pi)^3}\frac{d^3{\bf k}'}{(2\pi)^3} \frac{1}{({\bf k}-{\bf k}')^2+m_\rho^2}
\,\sum_N
\bigg(({\bf k}-{\bf k}')^2+3({\bf k}-{\bf k}')^2\frac{M^*}{E^*}\frac{M^{'*}}{E^{'*}}\nonumber\\
& &\qquad -({\bf k}-{\bf k}')^2\frac{{\bf k}^*}{E^*}\cdot\frac{{\bf k}^{'*}}{E^{'*}}+4({\bf k}-{\bf k}')\cdot\frac{{\bf k}^*}{E^*}({\bf k}-{\bf k}')\cdot\frac{{\bf k}^{'*}}{E^{'*}}\bigg)_N
\nonumber \\
& & +2\left(\frac{f_\rho}{2M_N}\right)^2 \int\frac{d^3{\bf k}}{(2\pi)^3}\frac{d^3{\bf k}'}{(2\pi)^3} 
\frac{1}{({\bf k}-{\bf k}')^2+m_\rho^2}\,
\bigg(({\bf k}-{\bf k}')^2+3({\bf k}-{\bf k}')^2\frac{M_p^*}{E_p^*}\frac{M_n^{'*}}{E_n^{'*}}\nonumber\\
& &\qquad -({\bf k}-{\bf k}')^2\frac{{\bf k}_p^*}{E_p^*}\cdot\frac{{\bf k}_n^{'*}}{E_n^{'*}}+4({\bf k}-{\bf k}')\cdot\frac{{\bf k}_p^*}{E_p^*}({\bf k}-{\bf k}')\cdot\frac{{\bf k}_n^{'*}}{E_n^{'*}}\bigg) \\
\epsilon_{Fock}^{(\pi)} &=& \frac{1}{2}\left(\frac{f_\pi}{m_\pi}\right)^2 \int
\frac{d^3{\bf k}}{(2\pi)^3}\frac{d^3{\bf k}'}{(2\pi)^3} \frac{1}{({\bf k}-{\bf k}')^2+m_\pi^2}\,
\sum_N \bigg(({\bf k}-{\bf k}')^2+({\bf k}-{\bf k}')^2\frac{M^*}{E^*}\frac{M^{'*}}{E^{'*}}\nonumber\\
& &\qquad -({\bf k}-{\bf k}')^2\frac{{\bf k}^*}{E^*}\cdot\frac{{\bf k}^{'*}}{E^{'*}}+2({\bf k}-{\bf k}')\cdot\frac{{\bf k}^*}{E^*}({\bf k}-{\bf k}')\cdot\frac{{\bf k}^{'*}}{E^{'*}}\bigg)_N
\nonumber \\
& & +2\left(\frac{f_\pi}{m_\pi}\right)^2 \int\frac{d^3{\bf k}}{(2\pi)^3}\frac{d^3{\bf k}'}{(2\pi)^3}
\frac{1}{({\bf k}-{\bf k}')^2+m_\pi^2}
\bigg(({\bf k}-{\bf k}')^2+({\bf k}-{\bf k}')^2\frac{M_p^*}{E_p^*}\frac{M_n^{'*}}{E_n^{'*}}\nonumber\\
& &\qquad -({\bf k}-{\bf k}')^2\frac{{\bf k}_p^*}{E_p^*}\cdot\frac{{\bf k}_n^{'*}}{E_n^{'*}}+2({\bf k}-{\bf k}')\cdot\frac{{\bf k}_p^*}{E_p^*}({\bf k}-{\bf k}')\cdot\frac{{\bf k}_n^{'*}}{E_n^{'*}}\bigg)
\end{eqnarray}
\vfill\eject

\end{document}